# Survey on Unmanned Aerial Vehicle Networks: A Cyber Physical System Perspective

Haijun Wang, Haitao Zhao, Jiao Zhang, Dongtang Ma, Jiaxun Li, and Jibo Wei,

*Abstract*—Unmanned aerial vehicle (UAV) networks are playing an important role in various areas due to their agility and versatility, which has attracted significant attention from both the academia and industry in recent years. As an integration of the embedded systems with communication devices, computation capabilities and control modules, the UAV network could build a closed loop from data perceiving, information exchanging, decision making to the final execution, which tightly integrates the cyber processes into the physical devices. Therefore, the UAV network could be considered as a cyber physical system (CPS). Revealing the coupling effects among the three interacted components in this CPS system, i.e., communication, computation and control, is envisioned as the key to properly utilize all the available resources and hence improve the performance of the UAV networks. In this paper, we present a comprehensive survey on the UAV networks from a CPS perspective. Firstly, we respectively research the basics and advances with respect to the three CPS components in the UAV networks. Then we look inside to investigate how these components contribute to the system performance by classifying the UAV networks into three hierarchies, i.e., the cell level, the system level, and the system of system level. Further, the coupling effects among these CPS components are explicitly illustrated, which could be enlightening to deal with the challenges in each individual aspect. New research directions and open issues are discussed at the end of this survey. With this intensive literature review, we try to provide a novel insight into the state-of-the-art in the UAV networks.

*Index Terms*—Unmanned aerial vehicle (UAV), cyber physical system (CPS), communication, computation, control.

## I. INTRODUCTION

Due to the characteristics of agility, versatility, low cost and easy-to-deploy, unmanned aerial vehicles (UAVs) are playing an important role in both military and civilian areas. In military applications, the UAVs are envisioned as an indispensable part of the future battlefield. They can not only capture different kinds of information on a large scale in terms of time and space proactively (e.g., border surveillance, intelligence reconnaissance), but also assist other unmanned/manned combat platforms to complete dangerous missions [1], [2]. The UAVs also flourish with the increasing demands of civilian applications, including agricultural plant protection [3], search and rescue [4], [5], environment and natural disaster monitoring [6], [7], delivery of goods [8], communication relays [9], [10], aerial base stations [11]–[13], construction [14] and traffic surveillance [15].

H. Wang, H. Zhao, J. Zhang, D. Ma, J. Li and J. Wei are all with College of Electronic Science, National University of Defense Technology, Changsha, 410073, China (e-mail: haijunwang14, haitaozhao, zhangjiao16, dongtangma, lijiaxun, wjbhw@nudt.edu.cn).

In these applications, one emerging trend is that the UAVs work from individually to cooperatively by constituting a reliable network, since multiple UAVs can provide wider coverage, more flexibility and robustness through redundancy [16]. Accordingly, the challenges in a multi-UAV network are much more intractable compared with those for a single UAV. Therefore, researches on the multi-UAV networks have attracted significant attention in recent ten years. Thanks to the advances in wireless communication, high performance computation and flight control areas, the UAV networks have obtained more powerful capabilities with respect to the communication, computation and control.

There appears a strong tendency to integrally design the communication, computation and control modules towards intelligent UAV networks [17]. This is reasonable since these three cyber components play crucial roles in the UAV networks, and they will affect and benefit from each other in a coupling way. For example, when the communication link is unstable due to the shadow effect, the UAVs could alter their locations through flight control to pursue line of sight links, other than only adjusting the communication parameters in vain. Thus, a cross-disciplinary viewpoint may bring inspirations to deal with the puzzles in each single area.

### A. CPS can inspire the UAV networks

The UAV network operates by building a closed loop, which consists of the initial data perceiving, information exchanging, decision making and the final execution. From this perspective, the complex UAV network can be considered as a cyber physical system (CPS), which has attracted significant attention recently [45]. CPS achieves the tight coupling between the cyber domain and the physical domain by strictly embedding the cyber processes into the physical devices. Thus, the reliable, real-time and efficient monitoring, coordination and control to the physical entities can be conducted through the closed loop [46]. For example, in the UAV networks, the data sensed by the sensors originates from the physical world (i.e., its mission circumstance), and the final decisions, made by computation and conveyed by communication, are translated into instructions and eventually take effects on the physical world through the actuators. Besides, CPS has a broad definition, therefore the UAV networks consisting of a single UAV (the cell level), a UAV swarm (the system level) or multiple heterogeneous UAV swarms (the system of system level) all could be developed as CPS. And as the next generation of systems, CPS is expected to be a key method to implement the artificial intelligence [45]. Thus, integrating a CPS vision into the UAV networks is promising to significantly improve the performance of the whole system.

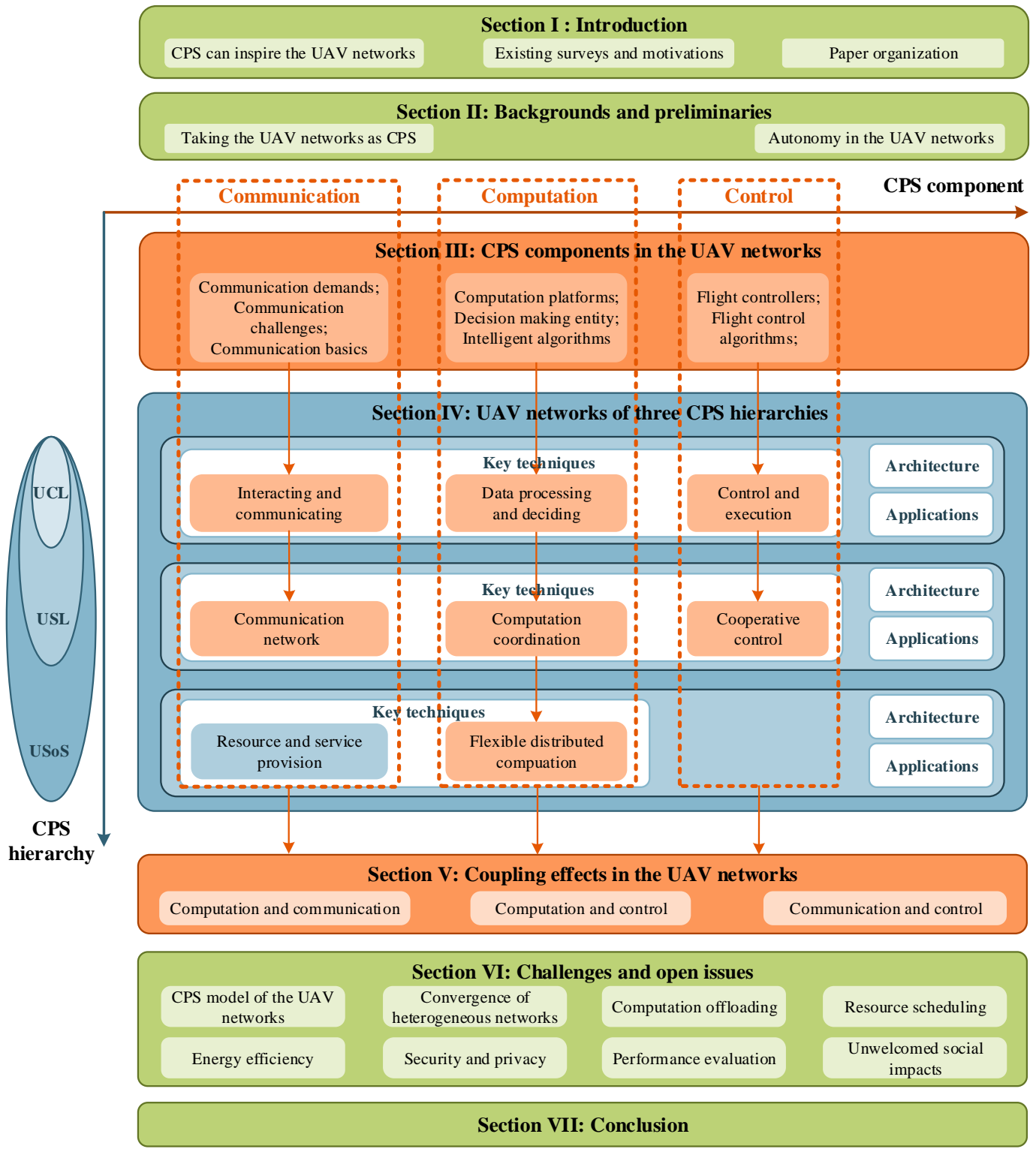

Fig. 1.  The structure of the survey.

TABLE I
A COMPARISON OF THE EXISTING SURVEYS ON THE UAV NETWORKS AND CPS

| Category | Surveys | Cyber issues | Physical issues | Coupling issue | UAV involved | Hierarchy | CPS applications |
|---|---|---|---|---|---|---|---|
| UAV network | [18], 2018 [19], 2017 [20], 2016 [8], 2016 [21], 2015 [22], 2013 | Communication and networking | Mobility, missions and energy | No | Yes | System | N/A |
| | [23], 2016 | Communication | Mobility and missions | No | Yes | Cell; System | N/A |
| | [24], 2017 [25], 2016 [26], 2015 | Computation and control | Autopilot platforms | No | Yes | Cell | N/A |
| | [27], 2018 [28], 2015 [29], 2012 | Flight control algorithms | Mobility | No | Yes | Cell | N/A |
| | [30], 2017 [31], 2013 | Formation control algorithms | Mobility | No | Yes | System | N/A |
| | [32], 2014 | None | Mobility model | No | Yes | Cell | N/A |
| | [33], 2013 | None | Collective mobility model | No | Yes | System | N/A |
| CPS | [34], 2014 | Communication | Communication infrastructure and physical dynamics | Yes | No | System | Smart grid |
| | [35], 2016 | Networking and security | Wind turbine components | Yes | No | Cell; System | Smart grid |
| | [36], 2016 | Security, attacks and defenses | None | No | No | System | Smart grid |
| | [37], 2016 | Security and safety | None | No | No | System | Medical devices |
| | [38], 2016 | Networking and protocols | Traffic flow and mobility models | Yes | No | System | Transportation |
| | [39], 2017 [40], 2016 | Security and threat detectors | None | No | No | System | General |
| | [41], 2017 | Security, privacy and defenses | None | No | Yes | System | General |
| UAV-enabled CPS | [42], 2015 | None | UAVs as sensors | No | Yes | System | General |
| | [43], 2014 | Contour mapping algorithm, flight and formation control | Fixed-wing and multirotor UAVs, mobility and missions | No | Yes | System | Source seeking and contour mapping |
| | [44], 2018 | Communication network, flight and formation control, and image analysis | Mobility, energy and UAV testbeds | No | Yes | Cell; System | General |
| *This work* | | Communication, computation and flight/formation control | Autopilot platforms, computation chips, mobility and missions | Yes | Yes | Cell; System; System of system | General |

### B. Existing surveys and motivations

The UAV networks and CPS have been extensively surveyed in many literatures. However, existing works do not well reveal the guidance of the CPS to the UAV networks.

On one hand, the existing surveys on the UAV networks are usually conducted from a single perspective, i.e., considering only the cyber issues or the physical issues. For example, in [8], [18]–[23], [47], the authors report the important concerns (including the challenges, characteristics, requirements and solutions) of UAVs all from a communication and networking viewpoint. [24]–[26] have surveyed the popular computation platforms for the UAVs. The control-related matters, including the control algorithms for single UAV and the formation control and coordination for multiple cooperative UAVs, are respectively investigated in [27]–[29] and [30], [31]. From the physical issue perspective, mobility models are presented for single UAV in [32] and for collective UAVs in [33]. However, the interactions, maybe the inspirations, between the cyber and physical domain as well as among the three cyber components, are not well excavated in these surveys, although they can be instructive to deal with the problems from a cross-disciplinary perspective. On the other hand, most existing surveys on the CPS put particular emphasis on the design and implementation of the industrial systems, concerning energy [34], [35], [48], transportation [38] and production [49]. Others focus on reviewing the security challenges and approaches in various CPS applications [36], [37], [39]–[41]. However, none of them have taken UAVs into considerations.

There are also some surveys combining the UAVs with CPS. In [42] and [43], the authors utilize the UAVs as sensors and actuators to implement CPS systems for various missions, such as cooperative target seeking and contour mapping. However, they only put emphasis on the specialized applications and implementations of the CPS, but ignoring the cyber issues and their inherent relations. In [44], the design issues and challenges with respect to the communication (i.e., networking and cross-layer design for scalable and secure communication), computation (i.e., image analysis and vision-based techniques) and control (i.e., flight control and path/trajectory planning) are respectively investigated for various UAV-enabled CPS applications. Nevertheless, the coupling effects between these components are not discussed.

Table I compares the existing review works, which indicates an imperative need for a comprehensive survey on the UAV networks from a CPS perspective.

### C. Paper organization

In this paper, we intend to provide a systematic review of the UAV networks from a CPS perspective. The main body of the paper is structured in a two-dimensional roadmap, i.e., the three cyber components as well as their coupling effects, and the hierarchies of the UAV networks, as shown in Fig. 1. We firstly give some backgrounds and preliminaries in Section II. Then the basic issues and advances with respect to the communication, computation and control of the UAV networks, are investigated in Section III. In Section IV, we categorize the UAV networks into three hierarchies according to the CPS scales, i.e. the cell level, the system level and the system of system level, and extract the key techniques in terms of the three components to build a UAV network under each hierarchy. In Section V, we further excavate the coupling effects and the cross-disciplinary issues between the cyber domain and the physical domain, as well as inside the cyber domain. In Section VI, we discuss some challenges and open issues for the future researches. Section VII concludes the survey.

To the best of our knowledge, this is the first survey that comprehensively investigates the key issues of the UAV networks from a CPS perspective. It is envisioned to provide a novel insight into the existing UAV researches, as well as some valuable suggestions to solve the challenges in this promising field.

## II. BACKGROUNDS AND PRELIMINARIES

### A. Taking the UAV networks as CPS

CPS is initially proposed by National Aeronautics and Space Administration (NASA) in space exploration which involves the unmanned aircrafts. Then it is extended to military applications to reduce the casualties, where the soldiers just need to sit in a command post and remotely control the weapons without being involved on the front. And now, it has been widely adopted in many society-critical domains towards "Industry 4.0", including the transportation, energy, healthcare and manufacturing [39], [50]. CPS can be regarded as the extension of the control systems and embedded systems [46]. And it flourishes with the rapid development of new information technologies, such as cloud computing, innovative sensing, communication and intelligent control [50]. The inherent idea of CPS is to monitor and control the physical world by integrating the cyber processes, including sensing, communication, computation and control, into the physical devices. Fig. 2 gives a general representation of the CPS in various applications.

As embedded devices, UAVs are equipped with onboard computers which are responsible for data processing, in other words, computing. The data, of course, derives from the sensing processes of various sensors, which includes the internal statuses of the UAVs themselves and the captured data from the outside world. Based on the information, the UAV network could make decisions about the flight control and the specific missions. And finally, these decisions take effects on the UAVs as well as the outside world through the control processes of the actuators (e.g., motors, rotors, wings and mechanical arms). Further, the UAV network could evaluate the effectiveness of the actions (also through the sensing and computation processes), to adjust the operation of the next cycle. In addition, the UAVs are also capable of exchanging the information with their counterparts (if there exists) through the communication network before and/or after the decision making to achieve a consensus. We can see that, the UAV network builds a closed loop that involves both the cyber and physical domain issues, and therefore, it can be seen as a CPS.

Specifically, the UAV networks can be decomposed as four key elements: *i)* the hardware, belonging to the physical domain, which directly interact with the cyber and physical world by sensing and actuating, or provide computation and communication capabilities. It may include the sensors, actuators, computation chips, communication equipment and so on; *ii)* the software, belonging to the cyber domain, which is used to manipulate the hardware, as well as analyze the data and make decisions. It may include the embedded operating systems, application programs, functional algorithms and so on; *iii)* the wireless communication network, belonging to the cyber domain, which is responsible for exchanging and sharing information among the entities in the system. It may include the communication modules, standards and protocols (e.g., the medium access control (MAC) and routing protocol); *iv)* the cloud service platform, belonging to the cyber domain, which is a highly integrated, open and shared data service platform. It can be a cross-system, cross-platform and cross-domain information center which incorporates data distribution, storage, analysis and sharing. The details about the cloud service platform will be illustrated in Part C, Section IV. Fig. 3 gives an example of a UAV network from a cyber and physical integration perspective.

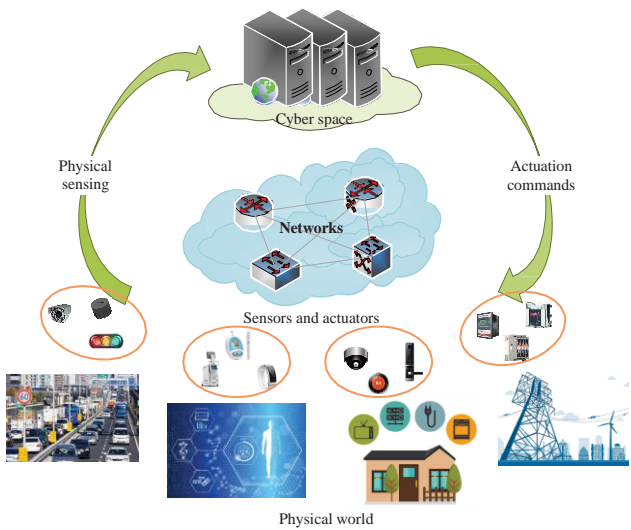

Fig. 2. A general CPS architecture.

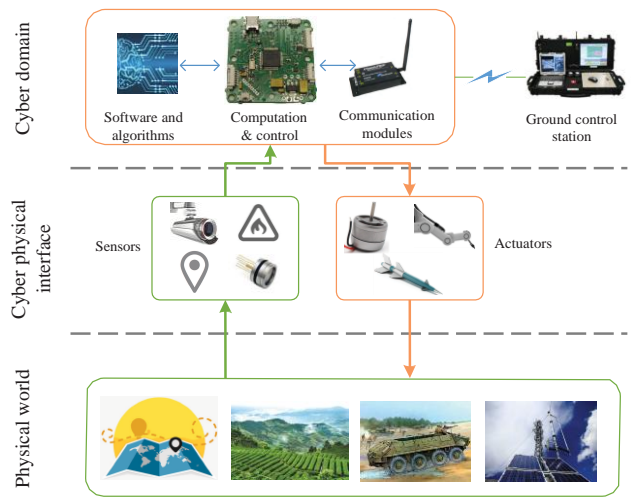

Fig. 3. An example of a UAV network from a cyber and physical coupling perspective.

TABLE II
Contribution comparison of the four elements to the UAV networks for different UAV autonomy levels

| UAV autonomy | Carrier | Three components of the UAV networks | | |
|---|---|---|---|---|
| | | *Communication* | *Computation* | *Control* |
| Human-in-the-loop | UAV | High | Low | High |
| | GCS | High | High | Low |
| Human-on-the-loop | UAV | Medium | Medium | High |
| | GCS | Medium | Medium | Low |
| Autonomous | UAV | High | High | High |
| | GCS | Low | Low | Low |

### B. Autonomy in the UAV networks

In most cases, one or multiple UAVs in the network are connected to a powerful ground control station (GCS), such as a human pilot or high-performance computers [47]. The GCS is used to monitor the statuses of the UAVs (e.g., the locations, health conditions), set waypoints and send out new commands. The control from the GCS to the UAVs is actually a computation-dominated process with the communication support, since that the GCS makes decisions based on the data transferred from the UAVs, and then sends the decisions to the UAVs, which are finally translated into instructions to control the UAVs. According to the intervention degree from the GCS, UAVs can operate in three autonomy modes [51]:

1) *Human-in-the-loop*: The UAVs could not operate independently and they are fully controlled by the GCS in real time. Thus, the flight control (e.g., waypoints, attitudes and navigation) and mission completion heavily rely on the skills of the human pilot, as well as the data link between the GCS and the UAVs. Under this mode, the GCS undertakes the computation tasks mainly, and the UAV network highly depends on the communication between the GCS and the UAVs;

2) *Human-on-the-loop*: It is also known as "semi-autonomous". The control operation is shared between

the GCS and an onboard flight controller. The onboard controller maintains the steady flight of the UAV, and makes flight adjustment to avoid collisions with the obstacles or its counterparts. At the same time, the GCS may assign commands to UAVs, such as path planning and task allocation decisions. In the system, the adjustable configuration determines the degree at which each of the two agents (i.e., the GCS and the onboard controller) contributes to controlling the aircraft. Under this mode, the computation tasks are decomposed to both the UAVs and the GCS. And the UAV network does not heavily rely on the communication between the UAVs and GCS since the semi-autonomous UAVs could still operate pretty well without a connection to the GCS for a while;

3) *Autonomous*: The UAV is capable of maintaining the optimal flight control, and making decisions about path adjustment and task decomposition automatically with the onboard computer, according to its operation environment. Under this mode, the UAVs could fulfill the computation tasks themselves. And the UAV network depends neither on the decisions of the GCS, nor the communication between the GCS and the UAVs, unless an emergency needs human intervention.

Table II lists the carriers (i.e., the UAVs and the GCS) of the three components (i.e., communication, computation and control) in system consisting of UAVs with different autonomy levels, and compares their respective contributions to the whole system. In this paper, we care more about the UAVs in autonomous mode.

### III. CPS COMPONENTS IN THE UAV NETWORKS

From the CPS perspective, a UAV network is an integration of sensing, communication, computation and control. They contribute to the closed loop together, to efficiently allocate the resources and optimally complete the missions. In detail, sensing introduces the original data into the UAV network from the physical world. Communication drives the data to flow inside of the UAV network, which guarantees the information distribution and sharing, and thus a global analyzing and deciding. Computation is the key of analyzing and decision making based on all the acquired information. And control focuses on translating the decisions into instructions through the actuators which act on the real world finally. Fig. 4 shows the relations and the data flow among these components in the UAV networks. Considering the sensing will be tightly associated with various sensors, and its focus may be biased to the operating principles, hardware composition and design, all these are not concerns of this survey. In this section, we will discuss some basics and advances about communication, computation and control of the UAV networks respectively.

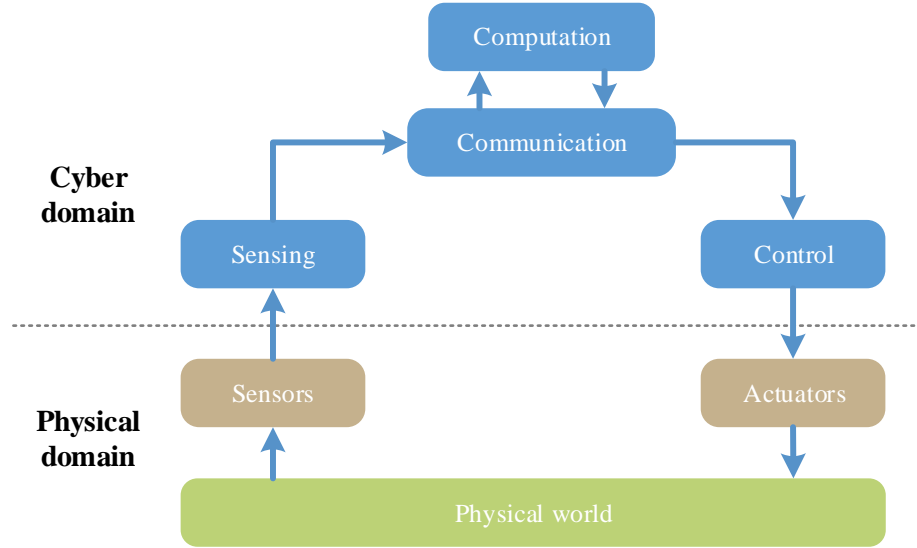

Fig. 4. The relations and data flow among sensing, communication, computation and control in CPS/UAV networks.

### A. Communication

Communication builds a "tunnel" for data from sensing to computation then to control in the UAV networks, as illustrated in Fig. 4. It guarantees the closed loop by driving the data flow considering that: *i)* the inputs of the computation, either in the GCS or in the UAVs, are obtained through communication; *ii)* the outputs of the computation, e.g., the decisions, are distributed (especially in the centralized deciding scenarios) or shared through communication (especially in the distributed deciding scenarios).

In the UAV networks, communication is not only imperative for disseminating observations, tasks and control information, but can assist in coordinating the UAVs more effectively and safely. The communication demands vary significantly in different applications. However, to realize robust communication is very challenging due to the intrinsic characteristics of UAVs, including the mobility and energy limitation, as well as the external constraints, including the spectrum scarcity and malicious interference. In the following text, we will first investigate the communication demands and challenges, and then discuss the basics and advances in communications.

(a) Communication demands

There are a variety of communication demands in terms of who will communicate and what to communicate. Generally, there are mainly two different communication types: *i)* in the UAV-GCS communication, the UAVs communicate with the GCS through the uplinks and the downlinks. The communication traffic between them may include the backhaul flight status and sensed data from the UAVs to the GCS, as well as the interventions/decisions from the GCS to the UAVs (e.g., the waypoints, flight control commands and mission plan); *ii)* in the UAV-UAV communication, any UAV can communicate with others directly or through multi-hop links. UAVs share the sensed data and decisions to guarantee a safe flight and a cooperative mission completion.

In both communication types, diverse messages need to be exchanged. From a more general perspective in regardless of the UAV autonomy, the transmitted traffic can be classified into three types as in [52], namely, control traffic, coordination traffic and sensed traffic. In detail, *i)* the control traffic exchange enables the GCS to monitor and influence the behaviors of the UAVs. It includes the mission commands and flight control messages from the GCS to UAVs. And also, the status data of the UAVs(e.g., the health status and telemetry data, including the inertial measurement unit (IMU) and global position system (GPS) information) is included, which is transmitted back to the GCS to provide a basis for decision making; *ii)* the coordination traffic means any data that needs to be exchanged for local decision making, cooperation and collision avoidance, without explicit input from the GCS. This kind of traffic may include the telemetry data, waypoint, mission plan and so on; *iii)* the sensed traffic encompasses the onboard sensor data used to measure the physical environment, which is transmitted to the GCS considering that onboard analysis of the sensor data may not be reasonable. It includes data transmission of various size (from weather sensor readings to high-quality images and videos) for real-time monitoring on the frontline, and decision making or post-mission analysis in the GCS. [52] lists task examples of each traffic and their corresponding quality of service (Qos) requirements, such as the delay, jitter and throughout.

The communication demands may change in the UAV networks with different autonomy levels and traffic types, according to the missions. First, for a network with UAVs in no or low autonomy, it depends highly on the UAV-GCS communications. However, for a network with UAVs in full autonomy, the UAV-UAV communications contribute much more to the system compared with the UAV-GCS communications, since that the UAVs just need to exchange information with their counterparts to make decisions rather than listening to the GCS. Second, different traffic types, including the control-related, mission-oriented and normal data messages, have various requirements to delay, delay variance and bandwidth [52]. Fig. 5 depicts the UAV network's communication dependency variance on the three kinds of traffics versus the autonomy level.

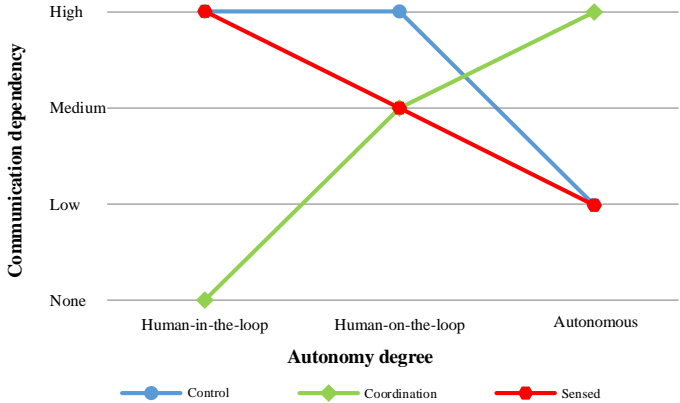

Fig. 5. The UAV network's communication dependency variance on the three kinds of traffics versus the autonomy level.

### (b) Communication challenges

The communication in the UAV networks faces many challenges, including the internal ones caused by the intrinsic characteristics of UAVs, e.g., mobility and energy constraint, and the external ones brought by the outside constraints, e.g., spectrum variation, scarcity and interferences.

*1) Internal challenges:* Mobility, in the form of wide operation space (sky in three-dimension), rich attitudes (pitch, roll and yaw) and varying speed (from static to high speed), is an inherent feature of the UAVs. Although static UAVs may be appropriate in some applications where the UAVs just need to hover in optimal locations (e.g., the aerial base stations and objective surveillance), in most cases, the mobility is more preferred since it expands the operating scope and reduces the number of UAVs deployed. However, the mobility will pose challenges over the communication of UAVs. First, the links are established intermittently since the UAVs may move with varying speeds depending on the missions. Second, the topology of the UAV network remains fluid with the changing link quality and the altering relative positions due to the mobility. Third, the Doppler effect cannot be neglected especially in the highly mobile scenarios, which will deteriorate the spectral efficiency and the communication performance [53]. Last but not least, the mobility brings about troubles to the antenna alignment, which needs to be considered when the directional antennas are equipped.

From another perspective, the mobility can be exploited well to improve the communication performance and network capacity by incorporating store-carry-and-forward (SCF) capabilities in the UAVs [54]. For example, in [23], the authors deploy a UAV to provide connectivity between two distant users, among which reliable direct communication links do not exist. The UAV flies back and forth between the source and destination to forward the data for both sides.

The energy of the UAVs is also constrained due to the limited payload capacity. On one hand, the communication distance and connectivity may be affected by the limited transmission power. On the other hand, the exhausting UAVs must fly back for charging and new ones may rejoin, which intensifies the topology changes. Interestingly enough, the authors in [54], [55] exactly exploit the exhausting UAVs and the returning time to improve the capacity of the UAV network. These UAVs could store and carry the data of other UAVs along their returning paths to the destination, achieving higher access efficiency and throughput.

So, we can summarize the internal challenges as the intermittent links, fluid topology, Doppler effect, complicated antenna alignment and vanishing nodes. Both of the survey papers, i.e., [8], [20], demonstrate the communication challenges caused by the high mobility and energy constraints of UAVs, and compare them with that in other two forms of ad hoc networks with mobility, i.e., mobile ad hoc network (MANET) and vehicular ad hoc network (VANET). They conclude that the communication technologies on the two mobile ad hoc networks could not well deal with the challenges resulting from the unique characteristics of the UAVs, including high mobility and limited energy.

*2) External challenges:* UAVs have been operating on the unlicensed spectrum bands, e.g., IEEE S-Band, IEEE L-Band, and Industrial, Scientific and Medical (ISM) band. And, most of them follow the policy of fixed spectrum assignment [53]. However, the UAVs still have problems on utilizing these spectrums, mainly because of the spectrum variations, spectrum scarcity and the outside interferences.

First, UAVs are deployed in diverse terrains according to the mission requirements, such as urban or suburb, in which the spectrum resources vary. Even in one mission, the UAVs may also suffer from temporal and spatial spectrum variation because of the changing operation area. Second, it is worthy to mention that there exist many other wireless networks in mission area generally, such as Cellular, Wi-Fi, Bluetooth and wireless sensor network. All these networks also work on the same spectrum band as the UAVs, which will cause interference and spectrum scarcity to the UAVs. Third, in a swarm with high UAV density, the UAVs also undergo the interference from their counterparts, and the malicious interference from the rivals, especially in the military scenarios.

Cognitive radio (CR) technology emerges as a promising solution to address the problem of spectrum scarcity [21]. It enables UAVs to exploit licensed or unlicensed spectrum bands opportunistically by using dynamic spectrum access techniques. Besides, the ability to adaptively change the working frequency also alleviate the effects caused by the spectrum variation and the outside interference. Various UAV applications, including traffic surveillance, crop monitoring, border patrolling, disaster management and wildfire monitoring, call for CR in communication and networking. However, many integration issues and challenges need to be addressed to make use of CR for UAV applications [8].

(c) Communication basics

When designing the communication of the UAV networks, there are several key basics that need to be considered, including the communication links, channel models, antenna designs and mobility models.

*1) Communication links:* In principle, there are two basic types of communication links in the UAV networks, i.e., the control and non-payload communications (CNPC) link, and the data link. The former is essential to ensure the safe operation of all UAVs in spite of the autonomy level, and the latter is to support the mission-related communications between the UAVs and the GCS as well as among the UAVs (mainly including the data transmission like images and videos). For the CNPC links, they long for much more stringent latency and security requirements to support safety-critical functions, including the command and control from the GCS to the UAVs, aircraft status report from the UAVs to the GCS and the sense-and-avoid information among the UAVs. So, CNPC links should operate in the protected spectrum in general, which have been allocated with two bands, i.e., the L-band (960-977 MHz) and the C-band (5030-5091 MHz) [56]. As for the data links, they mainly support for the direct mobile UAV-UAV communication and the UAV-GCS wireless backhaul. Compared with the CNPC links, the data links usually have higher tolerance in terms of latency and security requirements. Moreover, the UAV data links could reuse the existing band that has been assigned for the particular applications (by using CR technology). And also, dedicated new spectrum could be allocated to them for enhanced performance, e.g., millimeter-wave [57].

*2) Channel models:* In both the CNPC and data links, there are two types of channels, i.e., the UAV-ground channel (or A2G channel) and the UAV-UAV channel (or A2A channel). They are usually modelled differently due to their distinct channel characteristics. Besides, they have unique features compared with the extensively studied terrestrial communication channels [23].

For the UAV-ground channel, it is affected by the elements in the three-dimensional (3D) space which depends on the terrain the UAVs are flying over. For UAVs operating over the desert or sea, the two-ray model has been mostly used due to the dominance of the line of sight (LoS) link and surface reflection components [56]. However, the LoS links could be occasionally blocked by obstacles in most scenarios. In [58], [59], a much more common model is proposed. The authors consider the LoS link and the none line of sight (NLoS) link jointly. The path loss of the A2G channel is modelled as an average value of the LoS and NLoS components (both include the free space pathloss and an environment specific addition) under respective certain probabilities, which are mainly determined by the elevation angle and the environments, such as urban and suburban.

For the UAV-UAV channel, it is mainly dominated by the LoS component since that the UAVs always soar in the open sky [23]. Although the multipath fading exists due to the ground reflections, its impact is minimal compared to that experienced in the UAV-ground or ground-ground channels. In [60], a two-state Markov model is proposed to incorporate the effects of Rician fading depending on the distance changes between UAVs, which is suitable for strong LoS path for the UAV-UAV channels. Besides, the UAV-UAV channels may have even higher Doppler frequencies than the UAV-ground channel, due to the potentially large relative velocity between UAVs. Of course, Doppler effect can be handled well by adopting the frequency shift estimation or diversity technology, and there are lots of referring works, such as [61], [62].

Although, there are many researches concentrating on building the UAV-ground and UAV-UAV channel models, more further studies are needed to make the channel model more precise for different mission areas.

*3) Antenna designs:* The type and number of antennas that a UAV is equipped with need to be considered simultaneously. There are two types of antennas for UAV applications, i.e., the omnidirectional antenna and the directional antenna. The former radiates power in all directions, while the latter only sends the signal through a desired direction. Both of them have advantages and disadvantages, and they can also work together for a better performance. In brief, the omnidirectional antennas have natural advantages for UAV communication in highly mobile environments considering that they support the transmission and reception in all directions, but may result in safety issues, e.g., the communication leakage. The directional antenna usually have much longer transmission range than that of the omnidirectional antennas, which means less hop count and reduced latency between two UAVs. They could also improve the data security and handle the tradeoff between the communication range and the spatial reuse well [63], but may bring much more complicated protocols and harsh antenna alignment algorithms. [22] gives a brief comparison of the omnidirectional and directional antennas in UAV communications.

Besides, the number of the antennas also matters. On one hand, the advantages of the two kinds of antennas can be jointly exploited by converging multiple directional antennas to achieve omnidirectional coverage [64]. On the other hand, considering the traffic types between the UAVs, it is recommended that the UAV should be equipped with two types of antennas simultaneously. The omnidirectional one is used for exchanging the frequent information (e.g., the control and coordination information among the UAVs) without need to know the exact node location, and the directional one focuses on transferring the data with high rate requirement. In [65] and [66], the authors apply multiple-input-multiple-output (MIMO) into UAV communication to enable robust and high capacity connectivity between UAVs and the ground terminals, which needs to be supported by multiple antenna technology.

*4) Mobility models:* Considering the field tests are costly and with poor extensibility among different applications, designing suitable mobility models for the UAVs is of great significance to evaluate the communication related issues. Although, the traditional MANET models [67], such as random walk model, random waypoint model, random direction model and Gauss-Markov model, can be directly used or adapted for UAVs, they are incapable of capturing the correlation of aerial mobility for smooth turns. There are also a number of new models developed specifically for UAVs, including semi-random circular mobility model, three-way random and pheromone repel mobility model, smooth turn mobility model, flight-plan based mobility model and multi-tier mobility model. These models distinguish from the MANET models in that they capture smooth aerial turns caused by mechanical and aerodynamic constraint, and are suitable for different UAV applications. The authors in [32] have surveyed the mobility models for the airborne networks comprehensively.

TABLE III
A GENERAL TRAFFIC CLASSIFICATION AND THE RELATED BASICS IN THE COMMUNICATIONS OF THE UAV NETWORK

| Traffic type | Functions | Data | Size | Latency/security requirement | Entity | Link | Channels |
|---|---|---|---|---|---|---|---|
| Control | GCS monitors and intervenes the behaviors of the UAVs. | Mission commands; Flight control | Small | High | GCS to UAVs | CNPC link | UAV-ground; UAV-UAV |
| | | UAV status | Small | High | UAVs to GCS | CNPC link | UAV-ground; UAV-UAV |
| Coordination | For local decision making, cooperation and collision avoidance | UAV Status; Individual decisions (waypoint, mission plan) | Small | High | UAVs to UAVs | CNPC link | UAV-UAV |
| Sensed | Aerial imaging and measurements | Sensor data depending on onboard sensors and missions (videos, images, etc.) | Medium/big | Medium | UAVs to GCS | Data link | UAV-ground; UAV-UAV |

Table III lists the traffic classifications and the corresponding communication basics we have just talked about.

### B. Computation

Computation to the UAV network is just like the central nervous system to the human body system. In the closed loop, from the simple data analyzing to the complicated decision making, they all rely on the computation ability of the UAV networks. So, computation is the core of the closed loop. Besides, the information, decisions and its feedbacks from the physical world could be stored to form memories, knowledge and experiences. All these could further contribute to the decision making of the next cycle, which means that learning is also a part of the computation and will optimize the closed loop.

There are three basic issues that need to be discussed, including the onboard computation platforms, the computation/decision-making entities (i.e., who will make decisions, the GCS or the UAVs) and architectures (distributed or centralized), as well as the intelligent algorithms that could be used to optimize the decision making.

#### (a) Computation platforms

The computation abilities of UAVs derive from the computation platforms, which include the chip modules (mainly the main control unit (MCU)) and the corresponding embedded software (e.g., the operating systems (OS) and functional algorithms). For various application scenarios and autonomies, the UAVs may be equipped with different computation platforms.

*1) Hardware platform:* Existing UAVs for civilian applications usually adopt ARM-based MCU chips, such as the STM32 series of STMicroelectronics and Mega2560 series chips of Atmel [24]. These chips are characterized with low main frequency (e.g., the main frequency of the STM32 series is about 200MHz, and the Atmel's is as low as 20MHz), and poor computing power. They can only support the basic flight control for the UAVs with low autonomy, but cannot provide high-speed and parallel calculation capabilities for the UAVs which eager for intelligence.

Recently, the three chip giants, Qualcomm, Intel and Nvidia, all have advanced into this area and released their drone computation suites so as to promote the intelligence of the UAVs. In addition, the Chinese chip design company, Leadcore Technology, and the Chinese drone manufacturer, ZEROTECH Intelligence Technology, have jointly developed a solution for smart drones. They are Snapdragon Flight by Qualcomm, Edison for Arduino by Intel, Jetson TX1 by Nvidia (now updated to TX2 version) and LC1860 by Leadcore Technology respectively. Table IV lists the main parameters and characteristics of the four chip modules (where GPU for graphics processing unit, FLOPs for floating-point operations per second and CUDA for compute unified device architecture). They all have their own advantages in drone applications: *i)* the Qualcomm's module shines on its highest CPU frequency and the smallest size among all manufacturers; *ii)* for Intel's module, many indicators in Table IV are at a disadvantage. However, it can cooperate with its own environment sensor, i.e., RealSense technology, and thus has advantage of accuracy in environment perception over the binocular stereo vision adapted by other manufacturers; *iii)* Nvidia's module has the strongest floating-point parallel processing capability among the four, and it is capable of performing various types of image and pattern recognition, as well as advanced artificial intelligence tasks; *iv)* less information is currently available on the LC1860, and it is somewhat inferior to Qualcomm's and Nvidia's in some parameters, but it has a higher cost performance.

*2) Software platform:* In a UAV network, the software platform contains: *i)* the underlying firmware code, which connects the hardware systems with the software systems to ensure the efficient operation of the sensors, communication and computation units; *ii)* the functional software and algorithms, which may include the basic algorithms (e.g., the basic flight control, navigation and path planning algorithm), and some mission-oriented algorithms (e.g., the machine vision software for modelling the three-dimensional space of the UAV, and image and voice recognition algorithms for surveillance and reconnaissance missions). All of these can be enhanced by introducing the artificial intelligence algorithms, e.g., the deep learning [68], which in turn facilitate the usage and development of high-performance CPU on the UAVs; *iii)* the operating systems, such as Linux, Windows and Macintosh, which are necessary for running complex software and artificial intelligence algorithms on the smart hardware. Although traditional UAVs do not require a complicated operating system, many enormous and complex open-source projects, such as DroneCode, must rely on the operating system to manage the hardware abstraction interfaces and coordinate the computing resources.

TABLE IV
Comparison of the chip modules for smart UAVs provided by Qualcomm, Intel, Nvidia and Leadcore

| Manufacturer and module | Qualcomm Snapdragon Flight | Intel Edison for Arduino | Nvidia Jetson TX1 | Leadcore LC1860 |
|---|---|---|---|---|
| Advantages | Balanced performance of all aspects; Cost-effective | Adaption with the RealSense; High accuracy and extensive application range | Strong parallel calculation power | Low price |
| Disadvantages | Narrow application range | High power consumption and price | High price; Low communication performance | Low integrated technical performance |
| Size | 58mm*40mm | 127mm*72mm | 87mm*50mm | 41mm*61.5mm |
| CPU | 4*Qualcomm Krait 400 | 22nm dual-core and dual-threaded Intel Atom | 64-bit ARM A57 core | 6-core Cortex A7 |
| CPU performance | 2.5GHz | 500MHz | 2GHz | 2GHz |
| GPU | Qualcomm Adreno 330 | Intel HD Graphic | Maxwell architecture, 256 CUDA cores | Dual-core Mali T628 |
| GPU performance | 167GFLOPs | Unknown | 1TeraFLOPs | Unknown |
| Power consumption | Unknown | 35mW in static; Unknown in dynamic | Lower than 10W | Unknown |
| Wi-Fi & Bluetooth | Yes | Yes | Yes | Yes |
| Binocular Stereo Vision | Yes | RealSense support | Yes | Yes |
| High definition camera | 4K | Yes | 4K | 2K |
| Strong areas of the manufacturer | Baseband communication; Mobile computing | General-purpose computing; Advanced integrated circuit technology | GPU; Large scale parallel computing | Promising chip manufacturer in China |
| Strong areas for UAV applications | Low-power computing; UAV swarm communication | High performance computing | Machine vision computing; Artificial intelligence | Cost-effective |

There are many open-source software projects and libraries in the field of machine vision, artificial intelligence, including *TensorFlow*, *Torch*, *Caffe*, *OpenCV*, *CNTK*, *MXNeT* and so on. In the largest open-source software community, GitHub, a large number of machine vision and artificial intelligence algorithms can also be found. A brief description of these open-source projects is shown in Table V.

(b) Decision making entity

In general, all UAVs have computation ability to achieve the basic flight control, but only the UAVs with high autonomy level are capable of making decisions to fulfil the missions without interventions from the GCS. In the UAV networks, the decision making entities (DMEs) could be the UAVs or the GCS. Who will dominate relates to, also reflects the autonomy level of the UAVs. To be specific, the UAVs with low autonomy level must receive the decisions from the GCS all the time. However, the ones with higher autonomy level have much flexibility in making decision, i.e., by themselves in most time or by the GCS only in emergency.

Furthermore, the decisions can be made in a centralized or distributed manner. Utilizing a central DME offers a simpler solution than a distributed one, in terms of design and on-board processing power required of each UAV. However, the distributed decision making may be preferred to avoid a single point of failure, or increase time efficiency via parallel processing of multiple UAVs [8]. In this paper, the centralized decision making manner mainly refers to the GCS acting as DME to manage the UAVs with low autonomy level, and the distributed manner indicates the UAVs with higher autonomy level operating on their own.

The decision making manners also determine the communication demands, as illustrated in Part A, Section III. Even under the distributed manner, the communication demands may vary, depending on whether the decisions are made through consensus among the UAVs or only based on the individual status. Consensus-based decision making is expected to pose higher demands on the communication component than the individual-based one [69]. This is because of that the consensus-based decision making requires coordination information exchange among the UAVs while it is not necessary for the individual-based method.

Table VI summarizes the traffic to be exchanged for different decision making manners, where the telemetry information includes the IMU and GPS information. For the distributed decision making process, readers can refer to [70] for the individual-based method, and [71] [72] for the consensus-based method.

(c) Intelligent algorithms

Recently, there are many intelligent algorithms used to solve the mission-related decision making problems, such as path planning, task allocations, machine vision and image recognition. The intelligent algorithms include the bio-inspired intelligence algorithms (e.g., particle swarm optimization (PSO), ant colony optimization (ACO) and genetic algorithm (GA)) and the artificial intelligence related technologies (e.g., supervised learning, unsupervised learning, reinforcement learning and transfer learning). Of course, these intelligent algorithms are appropriate for different applications.

TABLE V
A BRIEF DESCRIPTION OF SOME OPEN-SOURCE SOFTWARE PROJECTS

| Project | Main area | Initiator | OS | Languages | Characteristics | Website |
|---------|-----------|-----------|-----|-----------|-----------------|---------|
| *TensorFlow* | Machine learning | Google | Linux; Windows; Mac | C++; Python | Support multiple platforms (CPUs, GPUs, TPUs, i.e., tensor processing unit) and devices (desktop, server, mobile, etc.); High-performance numerical calculation; Special optimization for multiple CPUs/GPUs | https://www.tensorflow.org/ |
| *Torch* | Machine learning | Facebook | Linux; Mac | Lua; C; Cuda | Use non-mainstream development language, Lua; Flexibility in implementing complex neural network topologies; Embeddable, with ports to iOS and Android backends | http://torch.ch/ |
| *Caffe* | Deep learning | UC Berkeley | Linux; Mac | C++; Cuda | Popular in image recognition and machine vision;Models and optimization are defined by configuration without hard-coding | http://caffe.berkeleyvision.org/ |
| *OpenCV* | Machine vision | Intel | Linux; Windows; Mac | C++; Python; Java; Matlab | Designed for computational efficiency and with a strong focus on real-time applications; Take advantage of the underlying heterogeneous computation platform | https://opencv.org/ |
| *CNTK* | Deep learning | Microsoft | Linux; Windows | C++; Python | Support both CPU and GPU devices; Parallelism with accuracy on multiple GPUs/machines; Easy to realize and combine popular models; Good at speech recognition | https://www.microsoft.com/en-us/cognitive-toolkit/ |
| *MXNeT* | Machine learning | DMLC/Baidu | Linux; Windows; Mac | C++; Python; Matlab | Emphasis on speeding up the development and deployment of large-scale deep neural networks; Easy to scale computation with multiple GPUs; Optimized Predefined Layers | https://mxnet.incubator.apache.org/ |
| *GitHub* | Software libraries | GitHub | Linux; Windows; Mac | Almost all | The largest host of source code in the world; Massive artificial intelligence related projects; The quality of the code varies greatly and they need secondary development for commercial use | https://github.com/ |

TABLE VI
THE DECISION MAKING ENTITY AND THE CORRESPONDING
COMMUNICATION DEMANDS

| Manner | | DME | Traffic exchanged | | Autonomy |
|--------|--|-----|-----|-----|----------|
| | | | *Min* | *Max* | |
| *Centralized* | | GCS | Telemetry; Control | Telemetry; Control; Sensed | Low |
| *Distributed* | Individual-based | UAVs | Telemetry | Telemetry; Sensed | High |
| | Consensus-based | UAVs | Telemetry; Coordination | Telemetry; Coordination; Sensed | High |

Bio-inspired intelligent algorithms are inspired by the biology activities in nature. For example, PSO imitates the foraging behaviors of the bird flocks or fish schools, i.e., birds or fishes forage individually and share information with others. ACO is enlightened by the behaviors of ants in their colony, which use pheromone as medium for communicating and creating traces for other individuals. Genetic algorithm is a metaheuristic inspired by the process of natural selection. It is commonly used to generate high-quality solutions for optimization and search problems, by relying on bio-inspired operators including mutation, crossover and selection. These swarm intelligence algorithms are enlightening to path planning and task allocation of UAVs considering the similarities between the UAVs and these organisms in achieving optimal collective behavior. In [73], the authors analyze how to combine swarm intelligence with multi-UAV task assignments after investigating the characteristics and principles of eleven swarm intelligence algorithms. The authors have employed PSO in [74], [75], ACO in [76], [77], GA in [1], [78] respectively to make real-time path planning for UAVs. In [79], the authors compare the parallel GA with PSO for real-time path planning and conclude that the GA produces superior trajectories compared with the PSO. The authors utilize PSO in [80], ACO in [81], GA in [82], [83] respectively to allocate tasks for the cooperative UAVs. And also, there are many modified and improved versions of these intelligent algorithms further optimizing the performance of the path planning [84], [85] and task allocation [86], [87].

The artificial intelligence algorithms can be used to improve the autonomy in making decisions, the recognition ability and adaptability to the dynamic environment of the UAVs. Thus, the dependency on the GCS-UAV communication performance can be reduced. The artificial intelligence algorithms can deal with the basic flight control issues [88], [89], path planning problems [90], [91] of the UAVs themselves, as well as some mission-related problems, such as the machine vision [92], pattern recognition [93] especially when executing reconnaissance and tracking missions. What the artificial intelligence algorithms bring to the UAVs is the ability to learn and utilize the experiences and knowledge, which makes the UAVs think and behave like a human. For example, the UAVs could recognize and classify what it saw by using (semi-)supervised learning [94] after being trained with considerable labeled data. However, when there is no enough prior knowledge, or it is too costly to label them, the unsupervised learning [95] is preferred. Sometimes, the UAVs need to interact with the environment so as to adjust their actions in real time and get more rewards, which can be achieved through reinforcement learning [96]. Transfer learning [97] is also beneficial to the UAV applications, which focuses on storing the gained knowledge and applying them to a different but related problem.

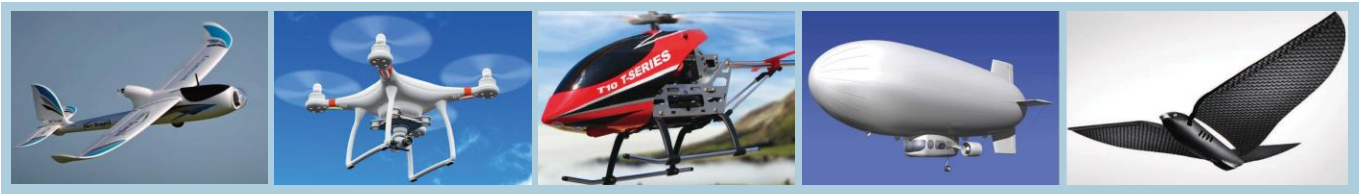

Fig. 6. Examples of different UAV forms and sizes: (from left to right) fixed-wing, multirotor, monorotor, airship and flapping wing.

*C. Control*

Control is responsible for the precise execution, which is embodied in a series of actions that affect the UAVs themselves and the physical world. The closed-loop data flow ends at control, and the previous sensing, communication and computation make sense only when the decisions are translated into instructions on the actuators and finally make a difference. In this section, we care more about the flight control of the UAVs, which is the precondition of the task execution.

The control to the UAVs varies with the UAV forms, such as the fixed-wing, multirotor, monorotor/single-rotor, airship and flapping wing, as shown in Fig. 6. They would suffer different aerodynamics, and are designated with different missions according to their characteristics [98]–[100]. For the specific missions, the control may include the macro flight control decisions (e.g., the path planning and formation control) and the micro executions of the actuators (e.g., the motors and rotors) according to the generated instructions. The path planning issues have been discussed in Part B, Section III based on the intelligent algorithms. The formation control/collective motion will be surveyed in the "key techniques" of Part B, Section IV, as well as in Part B, Section V, considering that they are the concerns of the UAV networks with multiple UAVs and usually bond with the computation techniques based on the swarm intelligence. Thus, in the following text, we will discuss the basic issues about the UAV control, including the flight controllers and flight control algorithms.

(a) Flight controllers

The flight controllers, also known as autopilot systems, are the core components of the UAVs. They are normally used to realize the autonomous flight control, including the attitude stabilization, fight waypoint generation, mission planning and so on [25]. In order to achieve these functions, the flight controllers need the hardware and software supports concurrently. The former mainly contains the onboard computers, inertial measurement units, various sensors, GPS modules, communication devices, power management modules and so on [26], and the latter usually includes the task allocation, path planning, waypoint generation, attitude control and signal processing algorithms [101]. The flight controllers enable the UAVs develop from the simple remote-controlled aircrafts to the fully autonomous and intelligent aircrafts. And these increasing requirements, in reverse, make the flight control platform to be more powerful, and meanwhile, more complicated [102].

There are many open-source communities on the fundamental flight controllers, including the hardware and software. In [24], [102], [103], the authors give an extensive survey on both the hardware and software of the open-source flight controllers, for small or micro UAVs systems in the market. In [104], the authors present a general view of the implementation of the open-source platform to develop a quadcopter research testbed, and survey a couple of open-source flight controllers. In the following text, well will investigate seven typical flight controllers respectively.

*1) Paparazzi:* It is the first open-source drone project, which encompasses the autopilot systems and ground station software for multi-copters/multirotor, fixed-wing, helicopters and hybrid aircraft [105]. Paparazzi features a dynamic flight plan system that is defined by mission states, and using waypoints as "variables". This enables UAVs to fulfil complex and fully automated missions without the operator's intervention.

*2) PIXHAWK:* Pixhawk is a high-performance and computer vision based autopilot-on-module, which is suitable for a variety of UAVs [106]. As an evolution of the PX4 flight controller system, it combines PX4-flight management unit (FMU) controller and PX4-IO into a single board. In addition, it works closely with the Linux Foundation DroneCode project. These properties make Pixhawk one of the most popular autopilots in the market.

*3) Phenix Pro:* It is built on the reconfigurable SoC (i.e., system on a chip). The flight controller is equipped with the real-time operating system (RTOS) and Linux-based robot operating system (ROS). Phenix Pro supports more than twenty interfaces, including the mmWave radar, thermal camera, ultra-vision HD video transceiver via software defined radio, etc. In addition, its hardware (FPGA, i.e., field programmable gate array) acceleration enables computer vision and deep neural network algorithm applications.

*4) OcPoC:* Octagnal Pilot on Chip (OcPoC) is developed by Aerotenna Company. Powered by the Xilinx Zynq processor and benefiting from FPGA superior performance, OcPoC possesses a greatly enhanced I/O capabilities and processing power. It runs ArduPilot software platform and implements real-time processing of sensor data simultaneously.

*5) DJI A2:* DJI develops a range of professional and amateur autopilot controllers for various multirotor platforms targeting at commercial and industrial aerial photography usage. DJI's autopilots are featured with reliability and robustness,

| Autopilot | Processor | Interfaces | OS | Redundancy | Control modes | Supported UAV forms | Open-source | Supporting software |
|---|---|---|---|---|---|---|---|---|
| *Paparazzi* | STM32F767 (ARM Cortex-M7) | UART, SPI, I2C, CAN, AUX | Linux | No | Auto; Assistant; Manual | Multirotor; Fixed-wing; Helicopter; Hybrid aircraft | Yes | Paparazzi UAV |
| *PIXHAWK* | STM32F427 (ARM Cortex-M4F) | SPI, I2C, CAN, PWM, ADC | Nuttx | IMU | Auto; Assistant; Manual | Multirotor; Fixed-wing; | Yes | DroneCode project |
| *Phenix Pro* | "Xilinx Zynq" FPGA SoC (ARM Cortex-A9) | CAN, HDMI, Camera Link, LVDS, BT1120-PL | RTOS, ROS | IMU | Auto; Assistant; Manual | Multirotor | Yes | RobSense Networked Robotics |
| *OcPoC* | "Xilinx Zynq" FPGA SoC (ARM Cortex-A9) | CAN, I2C, SPI, USB, UART, Ethernet | Linux | IMU | Auto; Assistant; Manual | Multirotor; Fixed-wing; | Yes | ArduPilot |
| *DJI A2* | Unknown | CAN | Unknown | Unknown | Auto; Assistant; Manual | Multirotor | No | Unknown |
| *NAVIO2* | Raspberry Pi 3 (Quad-core ARMv8) | UART, I2C, ADC, PWM | Linux (ROS) | IMU | Auto; Assistant; Manual | Multirotor; Fixed-wing; | Yes | ArduPilot |
| *Trinity* | Unknown | Unknown | Unknown | Autopilot system | Auto; Assistant; Manual | Multirotor | No | Unknown |

which contributes to the overwhelming market share of DJI UAVs. DJI A2 is not open-source, and the detailed information on its processor and sensor is not available either.

*6) NAVIO2:* As the latest version of the Navio autopilot family, Navio2 provides dual IMU chips to improve flight performance and redundancy compared to the previous versions. With the powerful Raspberry Pi board, Navio2 can not only realize the functions that Pixhawk has, but also provide more powerful computation to accomplish missions with higher levels.

*7) Trinity:* AscTec Trinity is the first fully adaptive control unit with up to three levels of redundancy for multirotor flight systems. Thus, it can automatically detect and compensate the system errors, and handle the sub-autopilot system failures. Compared with the aforementioned autopilots which only provide sub-module redundancy, such as IMU redundancy and power redundancy, Trinity can also provide total autopilot system redundancy, which makes Trinity more reliable, robust, and safe.

In Table VII, we compare the specifications of the seven flight controllers. Table VII and the analysis above indicate two development trends of the flight controllers. First, increasing the computation power and decreasing the energy consumption are two main methods to boost the performance of the UAV autopilots. Second, independent autopilot system redundancy is indispensable for the safe and robust control of the UAVs.

#### (b) Flight control algorithms

For UAVs of different forms, there are corresponding control algorithms aiming at guaranteeing their stable, smooth and safe flight, which encompasses the whole flight process from taking off, cruising to the final landing. The control of flapping-wing UAVs attracts only few researches because of their rareness and limited application scenarios [100]. The control of the fixed-wing UAVs or monorotor UAVs can be greatly enlightened by the traditional manned airplanes or helicopters [128], because of their similarities. As a new aircraft type, the multirotor UAVs, especially the multirotor UAVs, are attracting most research interests since they are competent in various scenarios. The unique flying abilities, such as vertical take-off and landing (VTOL), stable hovering and six freedom degrees, promote their extensive applications in military and civilian areas [99].

However, there are four challenges in designing the control algorithms for the quadrotor UAVs. First, as an intrinsic nonlinear system with static instability, it is difficult to build an accurate model for a quadrotor UAV. Second, the open-loop instability requires a fast control response and a large operation range [129] since the quadrotor UAVs are very sensitive to the external disturbances. Third, the quadrotor UAVs possess property of under-actuated, which results in strong coupling between the dynamic states. Fourth, some system parameters required in the control process, such as the inertial moments and aerodynamic coefficients, can hardly be measured or obtained accurately [130].

Surveys on the control algorithms for quadrotors to address the aforementioned challenges can be found in [28], [29], [131]. And all these existing algorithms can be broadly classified into two main categories:

- **Linear control**: Although the quadrotor is a highly coupled nonlinear system with multiple variables, it has been proved that most initial attempts to achieve autonomous quadrotor flight are based on the linear controllers, including the proportional integral derivative controller (PID), linear quadratic regulator/gaussian (LQR/LQG) and $H_\infty$. For trajectory tracking, the linear control can be applied only when the trajectory and the flying conditions for the quadrotor are not complex.
- **Nonlinear control**: It is developed to overcome the

TABLE VIII
The classification and comparison of the existing flight control algorithms for the quadrotor UAVs

| Category | Algorithm | Descriptions | Advantages | Disadvantages | References |
|---|---|---|---|---|---|
| Linear control | PID | It continuously calculates an error value as the difference between a desired setpoint and a measured process variable and applies a correction based on proportional, integral, and derivative terms. And it needs improvement to handle the uncertainties and external disturbances. | Simple structure; High stability and robustness; Efficient in hovering | Time-consuming; Incompetent for high disturbance | [107], [108] |
| | LQR/LQG | LQR describes linear system with the space state form, and its key design idea is to make quadratic objective function take minimum value by researching state feedback controller. LQG combines linear quadratic estimator with Kalman Filter for systems with Gaussian noise and incomplete state information. | Control several quadrotors; Simplify system in the scope of design allows | Only suitable for the linear system but not nonlinear systems except making some hypothesis | [109], [110] |
| | $H_\infty$ | It can deal with the problem of parametric uncertainties and external disturbances. Its improved version, $H_\infty$ loop forming, combines robust control thoughts with the classic loop forming, ensuring overall stability of closed-loop system. | No iterations when solving optimization problem; High robustness | Fail to control aircraft on a large scale | [111], [112] |
| Nonlinear control | SMC | It is based on Lyapunov stability criteria, and adjusts the error and its deviation of controlled plant to make controller move along the expected trajectory. It is designed by regulating the expectations Cartesian position and yaw angular velocity to make pitch and roll angle stability. | Robust to parameter variations and model uncertainties; Insensitivity to other disturbances | Easy to appear chattering phenomena | [113], [114] |
| | BC | It decomposes the whole controller into several steps and makes each step stable recursively. In each step, a virtual control is selected to make the former system stable. Thus, gradually modify control algorithm, until realizing regulation or tracking control, and ultimately gaining a relatively stable control effect. | Handle outside uncertainty well; Converge fast | Poor robustness, but can be compensated by other methods | [115]–[117] |
| | FL | It can algebraically convert (completely or partly) the nonlinear dynamic system into a linear system through adding state feedback, so that linear control techniques become applicable. It is also associated with other controllers, such as linear PID and $H_\infty$. | Good tracking performance compared to SMC; Flexible controller design | Sensitive to external disturbances, sensor noise and modeling uncertainty | [118]–[120] |
| | MPC | MPC uses an explicit dynamic model of the system to predict the future output behavior and minimize the tracking error over a future horizon by solving optimal control problems online. It is an advanced control technique, essentially a process of repeated optimizations and constraints at each time step. | Handle operational constraints explicitly than SMC, FL and BC | No computing time guarantee; Model accuracy dependent; High computational power | [121], [122] |
| | Adaptive control | It is a robust control method enabling system automatically adjust parameters to obtain the optimal control state during runtime, which is effective to reduce influence of parametric uncertainties and external disturbances. It usually consists of two closed-loop circuits: feedback loop circuit and parameter regulation loop circuit. | Cope well with uncertainties in mass, inertia matrix, and aerodynamic damping coefficients | All state feedback and exact model of predicted system are required, although it is difficult to get | [123], [124] |
| | Robust control | It guarantees controller performance within acceptable disturbance ranges or unmodeled system parameters. | Handle uncertainty in parameters or disturbances | Poor tracking ability. | [125], [126] |
| | NST | It is applied to the cascade nonlinear systems in strict feedforward form, which has little computation complexity, strong robustness and global stability results in the presence of control input saturations. | Ensure smoother UAV behavior; Less energy consumption | Less optimal, and high complexity | [127] |

limitations of the linear approaches, which are based on the nonlinear model of the quadrotor dynamics, including sliding mode control (SMC), backstepping control (BC), feedback linearization (FL), model predictive control (MPC), adaptive control, robust control and nested saturation technique (NST).

Table VIII classifies and compares the existing flight control algorithms for the quadrotor UAVs. The disadvantages of each control algorithm indicate the incompetence of single one to control the quadrotor UAV very well. Thus, combining two or multiple control algorithms jointly may bring better control performance. Actually, the linear control algorithms could cooperate with the nonlinear control algorithms. For example, a PID based sliding mode controller can be used to cope with the chattering problem under the circumstance of model error, parameter uncertainties and external disturbances [113]. A PID controller with feedback linearization and feedforward control, which uses backstepping method based on the simplified nonlinear dynamic model, can control the attitude and position of the quadrotor much better [132]. A controller which combines the feedback linearization method and $GH_\infty$ algorithm takes the best features of each [119]. In [133], the authors present a control method based on MPC and PID for path following of a quadrotor UAV to achieve rapid response for attitude, where the MPC controllers are designed to track the reference trajectory in the outer loop (position loop), while the PID controllers are designed to track the reference attitude in the inner loop (attitude loop). Also, the nonlinear control algorithms could complement with each other. For example, combining MPC with the robust feedback linearization method can deal with the leader-follower formation control problem of the quadrotors [134]. A sliding mode controller based on backstepping algorithm performs better on controlling the quadrotors [117].

## IV. UAV NETWORKS OF THREE CPS HIERARCHIES

The UAV networks are mission-oriented in most cases, and missions of different scales and complexities require the deployment of UAV networks with different hierarchies [8]. In detail, for a simple and low-scale mission, a UAV network comprised of a single UAV (associated with a GCS) can be appropriate. For more complicated missions in a larger scale, multiple interactive UAVs, i.e., a UAV swarm, may perform better through redundancy and swarm intelligence. Further, for a systematic, complex and spatial-temporal mission(e.g., anti-terrorists joint battle), a UAV network with several UAV swarms of different functions is preferred. These multiple UAV swarms can be connected to a cloud service platform, where they could share the information with others. Accordingly, the UAV networks can be divided into three hierarchies from a CPS perspective, i.e., *UAV network of cell level (UCL)*, *UAV network of system level (USL)* and *UAV network of system of system level (USoS)*.

UCL could run independently by building a local closed loop from self-sensing, self-analyzing, self-deciding to self-executing based on the onboard hardware and the embedded software. So, UCL can be seen as "hardware + software". USL could operate efficiently by building a larger-scale closed loop with the help of the communication network among multiple UAVs. Therefore, USL can be regarded as "hardware + software + communication network". USoS achieves the cross-platform and cross-system interoperability among multiple USLs or UCLs by constructing a cloud service platform. Further, a UAV industry ecology can be established by enriching the development tools, opening the application interfaces, sharing the data resources, building the development communities, and boosting all kinds of applications and software. Therefore, the USoS is an organic combination of all kinds of UAVs, and it can be treated as "hardware + software + communication network + platform". The hierarchy evolution of the UAV networks is illustrated in Fig. 7.

### A. Cell level

UCL is the minimum and basic unit, which cannot be partitioned into smaller CPS unit. A single UAV could be regarded as a CPS of the cell level considering itself can build a data-driven closed loop to optimize the resource allocation. The communicating ability is also indispensable, since the UAV will interact with the GCS, as well as with other UAVs when constructing a USL. Thus, the UCL is capable of perceiving, computing, communicating, controlling and extending. Under this hierarchy, the sensing, controlling and computation hardware, the embedded software and the communication modules define the functions of the UCL together.

Here, we will take the packet dispatching as an example, which is regarded as an effective method to solve the "last mile" problem of logistic. A UAV with a packet departs from the distribution station to the destination. The UAV initially makes a rough path planning according to the prior information. During the cruise, the UAV needs to collect the real-time data through various sensors onboard, including the outside information (e.g., navigation, altitude, air speed, barometric pressure, humidity and potential obstacles) and the inside information (e.g., barycenter variance, body humidity, energy consumption and the operation status of subsystems). Then the data is fused and analyzed by the embedded software. And the real-time flight decisions are made, which may include the path adjustment resulting from some unforeseen obstacles and heavy weather, or a sudden evasion response to an oncoming bird. At last, the control instructions on the actuators are generated according to the decisions, and will take effects. In a sudden situation, the UAV is controlled with an auxiliary decision made by the GCS according to the backhaul information.

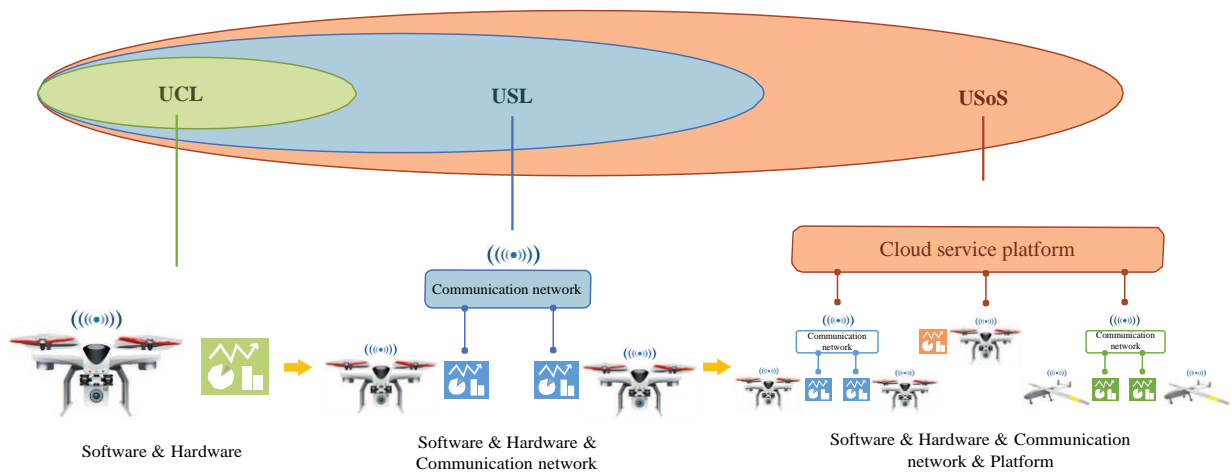

Fig. 7. The hierarchy evolution of the UAV networks.

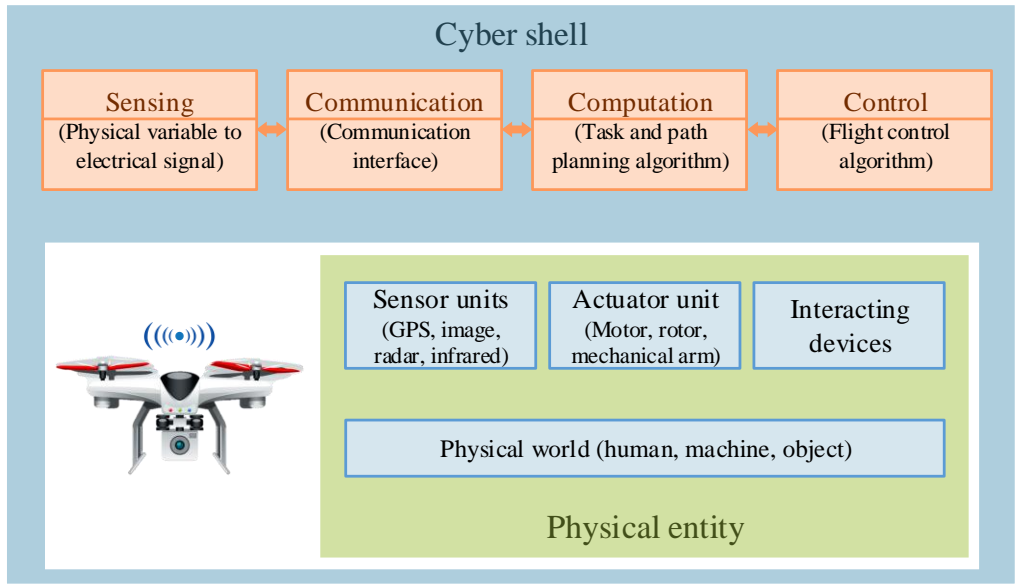

Fig. 8. The constitute architecture of the UCL.

**(a) Architecture**

*1) Constitute architecture:* The physical entities and the cyber shell constitute the UCL together, as illustrated in Fig. 8. In detail, *i)* the physical entities mainly include the physical world, i.e., the operation environment (e.g., the humans, machines and objects) and the onboard devices (e.g., sensors, actuators and other interacting devices) which interact with the outside environment. The physical entities are the manipulation parts of the physical process. First, they can sense and monitor the external signals, physical conditions (e.g., light, heat) or chemical composition (e.g., smog) through sensors. Second, they can receive the control commands and further exert control effects on the physical world through the actuators; *ii)* the cyber shell mainly includes the functionalities such as sensing, computation, control and communication, which is the interface between the physical entities and the information world.

The physical entities realize the digitalization through the cyber shell. The cyber shell builds a bridge for the physical entities to exchange information with the outside. Together, information world controlling the physical entities can be achieved. Therefore, the physical domain and the cyber domain are bound together.

*2) Organization architecture:* The organization architecture (or the network topology) of the UCL is usually simple. It is usually conducted with an infrastructure-based topology, where the GCS works as the center. The UAV connects with the GCS directly and receives the intervention from the ground. Even in some cases where multiple UAVs are preferred, the UAVs will neither communicate with each other directly nor with the help of the GCS as a relay. The GCS controls each UAV respectively. Fig. 9 gives two examples of the topology of the UCL.

**(b) Key techniques**

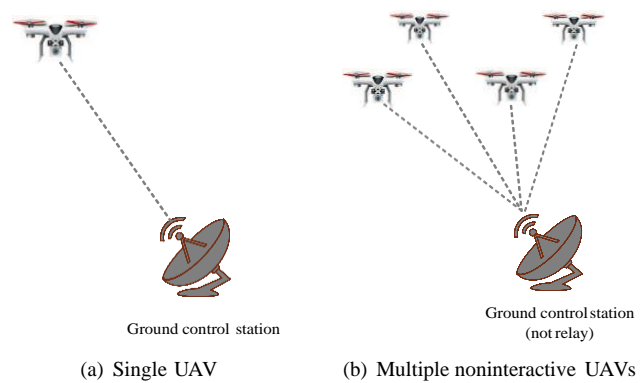

| (a) Single UAV | (b) Multiple noninteractive UAVs |
|---|---|

Fig. 9. Two examples of the UCL topology.

The key techniques define the technical requirements that need to be considered when constructing a UCL. According to the UCL architecture, the sensors are the data sources to obtain relevant information for UAVs. The acquired data further needs to be analyzed, which will circulate in the cyber domain. The actuators implement the control to the physical world according to the computation result. Therefore, the technical requirements can be summarized: *i) data processing and deciding*: the raw data is transferred into information and knowledge, which further reflects the current and will predict the ongoing state of the UAV. Some data processing technologies can be adopted, such as data mining, machine learning and cluster analyzing; *ii) control and execution*: the UAV acts on the physical world according to the decisions, which guarantees the mission fulfilment; *iii) interacting and communicating*: the UAV will receive the intervention from the GCS in emergency. Also, the USL calls for the communication ability of the individuals to build a self-organized network. Fig. 10 shows an overview of the existing researches on the key

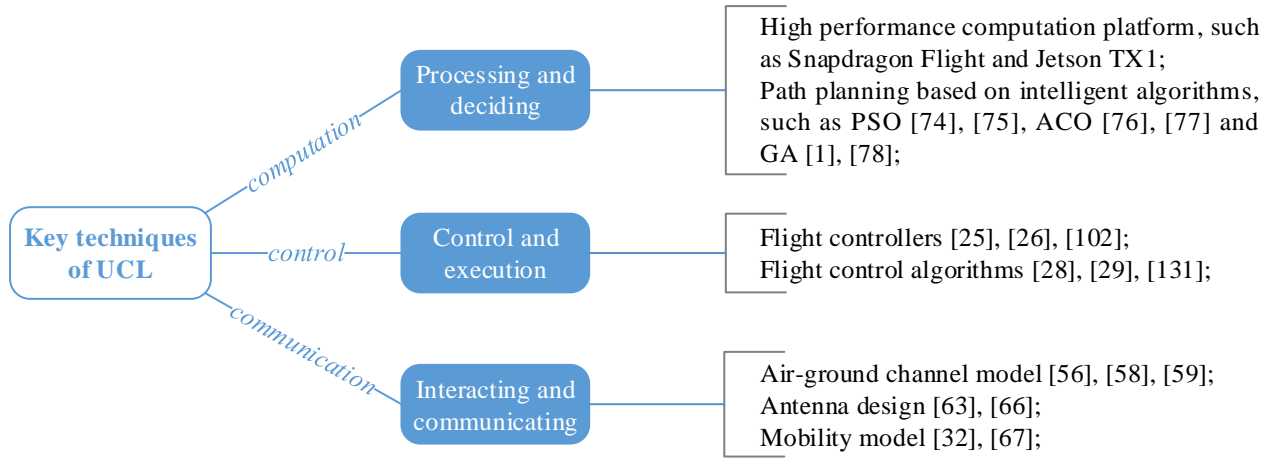

Fig. 10. The key techniques of UCL.

techniques of the UCL.

(c) Applications

Considering the agility and low cost of a single UAV, UCL is widely applied to various scenarios. The single UAV only needs to maintain the connectivity with the GCS at most, free from being imprisoned by the connections with other UAVs. The applications may include: *i)* UAV-aided relying, where the UAV flies back and forth to delivery data between two isolated users [135]; *ii)* communication coverage, where the UAV hovers as an aerial base station [12], [13], or flies to disseminate and collect data for the ground sensors [23]; *iii)* sensing and actuation coverage, where the UAV makes low scale surveillance and reconnaissance for the crops, air pollution, forest fire or wild animals [3], [6]; *iv)* logistics, where a UAV heads for the predefined destinations and completes the delivery action autonomously, especially for the remote area [8].

B. System level

Based on UCL, the communication network could be introduced to achieve the efficient coordination and optimized resource allocation among the UAVs [47]. Under this hierarchy, multiple UAVs cooperate with each other from the spatial and time dimensions (e.g., several UAVs executing a large-scale reconnaissance mission), or from the function dimension (e.g., several UAVs with different but complementary functions doing a military operation). A closed loop of larger scale can be built and enhanced in USL, of which the differences compared with that in UCL lie in that: *i)* UAVs may communicate to share the information/decisions after data sensing and processing, which helps a lot in building a comprehensive overview of the whole system for each UAV; *ii)* the decision-making manner is more flexible, which can be centralized, distributed or hybrid.

(a) Architecture

*1) Constitute architecture:* The constitute architecture of the USL is illustrated in Fig. 11, where multiple UAVs operate cooperatively with the help of the communication network. The communication network enables the automatic data flow of wider range, i.e., from the self-loop (in the UCL) to the system-loop (in the USL). The communication network also helps each UAV to maintain a complete belief map of others through their observations and interactions [136], which facilitates the global decision making and the swarm intelligence. Therefore, the interoperability among the UAVs is achieved, and the depth and precision of the resource configuration are improved.

Except for the functions of the UCL, USL mainly acquires the abilities of mission coordination, joint path planning, cooperative control, monitoring and diagnosing, and data interoperating. For example, cooperative control guarantees the linkage and coordinated control of multiple UAVs. Monitoring and diagnosing are mainly to monitor and diagnose the status of each UAV in real time. All these functions and decisions are supported by the communication network among them.

*2) Organization architecture:* Compared with UCL, the organization architecture of USL is more diverse and flexible. In general, the topology of the USL can be constructed as:

- **Infrastructure-based**: As shown in Fig. 12 (a), all UAVs could connect with the GCS directly and the communication among them will be relayed by the GCS. This may result in link blockage, higher latency and large-bandwidth downlink requirement [20]. As for the decision making entity, the GCS tends to take charge of the whole system, considering its acquisition of the global information. Of course, the UAVs can also make their distributed decisions after obtaining the information through the infrastructure-based communication.

- **Star**: Under this architecture, all UAVs would be linked to a central UAV directly, which is designated in advance or temporarily selected by other UAVs. All the communications among UAVs would be routed through the central node. Therefore, the central UAV may become the bottleneck of the system, which also causes link blockage and high latency. There may also exist multi-star topology, where the UAVs would form multiple clusters and each cluster is organized as a star. Both in single-star and multi-star architecture, the central nodes will connect with the GCS directly. Considering the UAVs could communicate with each other with the help of the central UAVs, other than through the GCS only, the decisions would be made by UAVs themselves in a distributed individual-based or consensus-based manner. Fig. 12 (b) and Fig. 12 (c) show the single-star and multi-star configurations respectively.

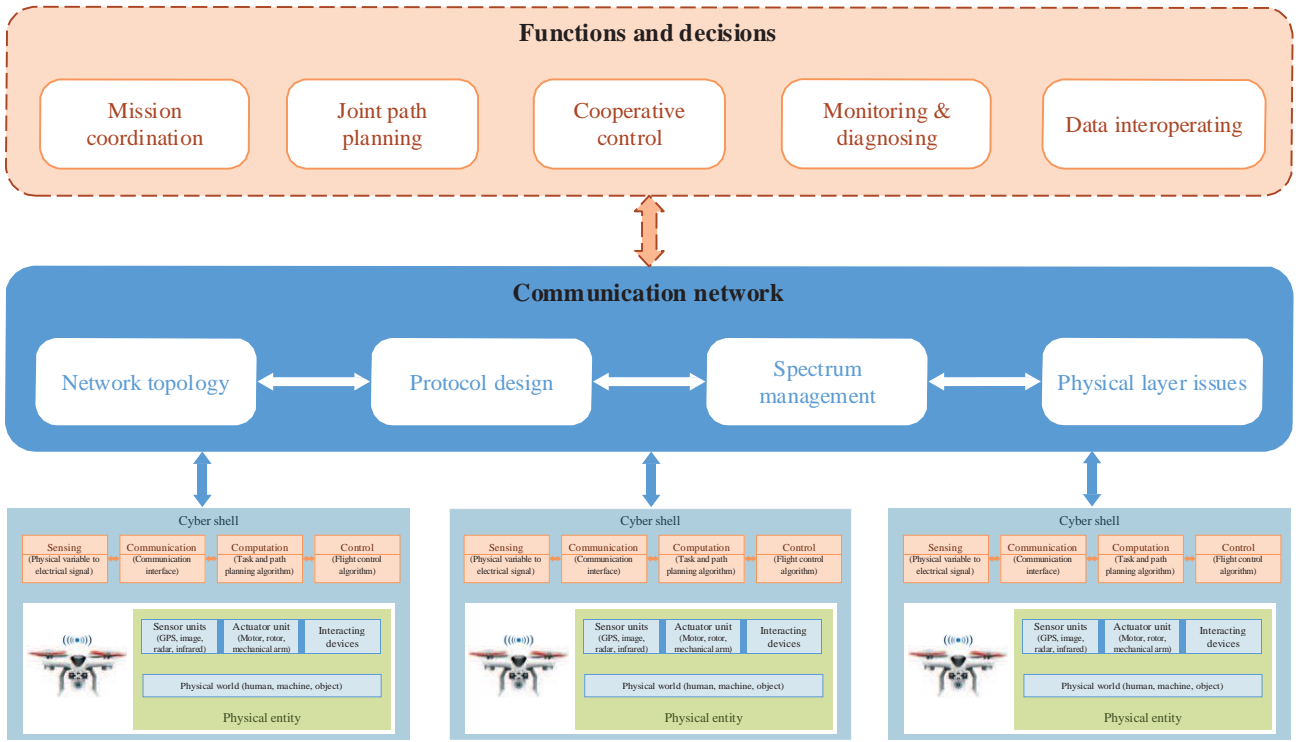

Fig. 11. The constitute architecture of the USL.

- *Ad hoc*: The UAVs could communicate with each other through the single-hop or multi-hop link without a central node. Single node failure has little effect on the whole system. This architecture would bring reduced downlink bandwidth requirement and improved latency because of shorter links among UAVs. Only one or several UAVs need to connect with the GCS, so that the coverage area of the network is significantly extended. Similarly, the UAVs also can be organized as several clusters, and each of them is configured as an ad hoc network. We call this hierarchical ad hoc architecture. Under this architecture, the UAVs would make distributed decisions by themselves. Fig. 12 (d) and Fig. 12 (e) give the examples of the flat and hierarchical ad hoc networks respectively.

For USL, all the three kinds of organization architectures have their own characteristics. They can be applied to different scenarios according to the missions, as well as the amount and performance of the UAVs. Actually, the three kinds of architectures could transform to each other if necessary. The authors in [137], [138] have studied the architecture self-adaption algorithm, which improves the performance of the communication network by supporting the transformation among three architectures adaptively.

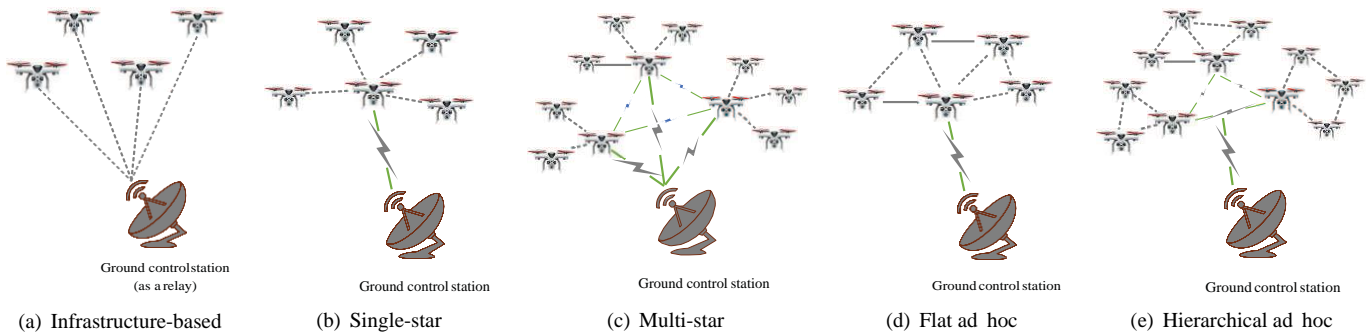

Fig. 12. Five representative examples of the USL topology.

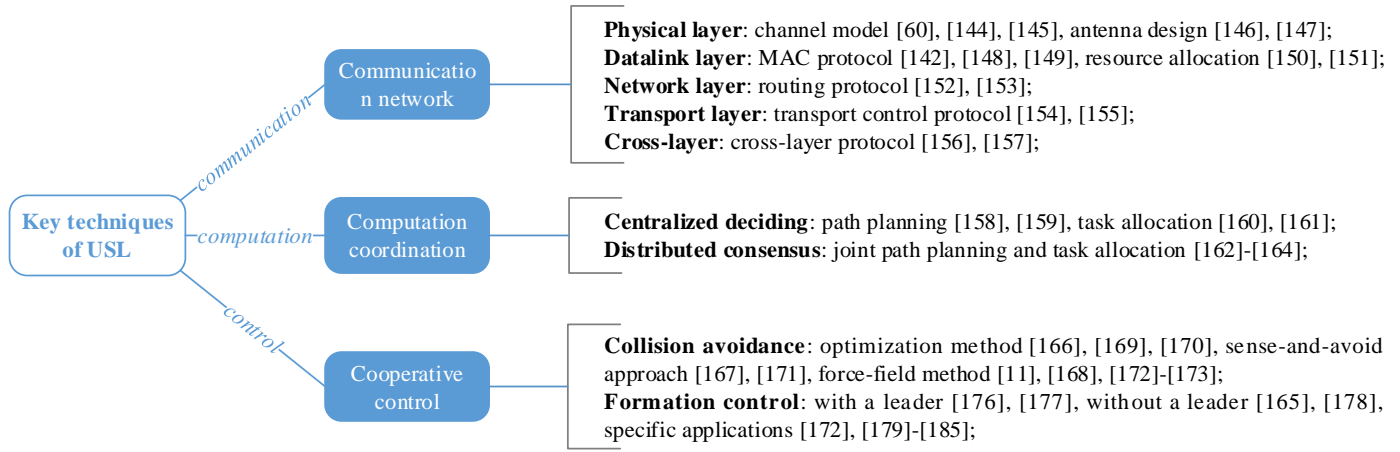

Fig. 13. The key techniques of USL.

**(b) Key techniques**

Referring to the constitute and organization architecture, the interconnections among the UAVs are mainly concerned in USL. The massive computation resources, which are contributed by all the individuals, need to be coordinated well when selecting the DME and assigning the computation tasks. USL also focuses on the real-time and dynamic control to each UAV, which realizes the unification of the cyber and physical domain. Therefore, it can be concluded that, based on the technical requirements of the UCL, the following issues need consideration when constructing a USL: *i)* the communication network; *ii)* the computation coordination, i.e., where and how the computation tasks are undertaken (e.g., in the UAVs or the GCS, with a centralized or distributed manner, as illustrated in Part B, Section III); and *iii)* the cooperative control. Fig. 13 shows an overview of the existing researches on the key techniques of USL.

*1) Communication network:* Communication network enhances the closed loop considering that the UAVs benefit a lot from the shared information from their counterparts when making the optimal decisions. However, it also brings challenges to the closed loop because the communication network can be a bottleneck of the whole system.

Although a set of off-the-shelf candidate technologies can be exploited for the communication networking of the USL, including IEEE 802.11 [139], IEEE 802.15.4 [140], 3G/LTE, and infrared [141], a widely accepted communication networking solution further needs to be studied [52]. It is expected to handle the harsh environment of the aerial links and be applicable to a broad scenarios. Considering the mobility and power limitation of the UAVs, the communication network should be characterized with low latency for the real-time control information exchanging, robust to deal with the vanishing links/missing nodes/fluid topology, high data rate for multimedia transmission and so on. And there are many researches heading for achieving all these requirements from different perspectives when designing the UAV communication network, including [52], [142].

The communication network design is usually conducted from a layered perspective, mainly including the physical layer basics (e.g., the radio propagation modelling, antenna architectures and spectrum band selection, which can be found in Part A, Section III), data link layer design (e.g., the medium access control (MAC) protocol, channel allocation and rendezvous [143]), network layer design (e.g., the route selection and Qos guarantee [18]), transport layer design (e.g., the congestion control and flow control) and some cross-layer considerations. Actually, the issues in each separate layer will have impact on others, and they have to be tuned well for a satisfied network performance. There are many surveys focusing on dissecting the important issues and giving some design instructions on the UAV networks, including [8], [20], [22], [47]. Table IX gives an overview of the researches on the UAV communication networks from a layered perspective.

*2) Computation coordination:* In USL, the computation resources are dispersed to separate entities, including the GCS and the UAVs. So, the computation resources need to be coordinated well, i.e., various computation tasks should be designated to the proper DMEs at the right time. However, where and how the decisions are made is an open problem, which can be correlated with many issues.

First, computation coordination depends on the computation task requirements. For example, the basic flight attitude control (e.g., pitch, roll and yaw) of each UAV could be guaranteed by the onboard controller [98], while the complicated and computation-intensive tasks (e.g., the image recognition in a reconnaissance mission) should be transferred to the GCS, since that the latter possesses more powerful computation ability. For other moderate computation tasks, such as the path planning, mission allocation and so on, the computation entities and decision making manners are much more flexible. They can be undertaken only by the GCS or the UAVs, or both together.

Second, computation coordination is also related to the network topology. In detail, under the infrastructure-based topology, the decisions for each UAVs tend to be made and optimized by the GCS in a centralized manner, since it could obtain the global information of the system. [158], [159] and [160] respectively optimize the path planning and task allocation problem both by employing a centralized strategy. In [161], the authors have addressed the task assignment and trajectory planning subproblems concurrently using the genetic algorithm and Voronoi diagram with a centralized approach. However, in the USL with star or ad hoc topology, the UAVs themselves will participate more in making decisions based on the exchanged information with each other, other than totally relaying on the GCS since that: i) under the two topologies, the multi-hop link of the UAV-GCS communication may bring increased latency, which will deteriorate the performance of the whole system; ii) the GCS should not be the fatal point of the system, i.e., the GCS failure should not paralyze the USL totally.

TABLE IX
AN OVERVIEW OF THE RESEARCHES ON THE UAV COMMUNICATION NETWORKS FROM A LAYERED PERSPECTIVE

| Layer | Problem | Researches | Brief descriptions | Applicable topologies |
|---|---|---|---|---|
| Physical layer | Channel modelling | A2G channel modelling [58] | The pathloss is modelled as an average value of the LoS and NLoS components under certain probabilities which depend on the elevation angle and the environments. | Infrastructure-based |
| | | A2A channel modelling [60] | A two-state Markov model is proposed to incorporate the effects of Rician fading depending on the changes of the distance between UAVs. | Star; Ad hoc |
| | | General link outage model [144] | A general analytical formula is provided for the outage of A2A and A2G links over various fading channels. Rayleigh, Nakagami-m, and Weibull models are studied as fading channels. | Infrastructure-based; Star; Ad hoc |
| | | MIMO A2G channel modeling [145] | The ground surface and roadside environment reflections are considered to investigate the statistical properties of the proposed 3D elliptic-cylinder UAV-MIMO channel model. | Infrastructure-based |
| | Antenna architecture | Antenna structures and types [146] | UAV receivers can achieve poor packet reception correlation at short time scales. The usage of multiple transmitters and receivers improves packet delivery rates dramatically. | Infrastructure-based |
| | | Multiband segmented loop antenna [147] | The antenna is composed of a segmented loop including eight segments, a patch element, and a shorting strip. It operates with an omnidirectional radiation at 956 MHz. | Infrastructure-based; Star; Ad hoc |
| Data link layer | MAC protocol | Position-prediction-based directional MAC [142] | The protocol includes position prediction, communication control, and data transmission phases. The exact position of each node is estimated at the position prediction phases. | Ad hoc |
| | | Multiple access scheme for super dense aerial sensor networks [148] | It is a combination of CSMA/CA (i.e., carrier sense multiple access with collision avoidance) and TDMA (i.e., time division multiple access) protocols based on a hybrid collision coordination technique. A spatial diversity technique and a simultaneous data transmission scheme based on a partnership of two nearby UAVs are introduced. | Star |
| | | Location oriented directional MAC protocol [149] | It uses a directional antenna to overcome the problems under an omnidirectional antenna as well as the well-known deafness problem of directional MAC by using busy-to-send packet along with request-to-send and clear-to-send packet. | Ad hoc |
| | Resource Allocation | Resource allocation under limited bandwidth [150] | It can allocate the resource to satisfy the high network throughput as well as the minimum data requirement in the given network environment. | Infrastructure-based; Star; Ad hoc |
| | | Resource allocation for packet delay minimization [151] | It could minimize mean packet transmission delay in 3D cellular network with multi-layer UAVs, which was obtained in M/G/1 queue formulated in UAV. | Star; Ad hoc |
| Network layer | Routing protocol | A reinforcement learning based routing protocol [142] | The routing process is regarded as a partially observable Markov decision process, which provides an automatically evolving and more effective routing scheme. | Ad hoc |
| | | Robust and reliable predictive routing strategy [152] | The scheme features hybrid use of unicasting and geocasting routing using location and trajectory information. It increases the robustness and reliability of the established routing path. | Ad hoc |
| | | Predictive optimized link-state routing (OLSR) [153] | It takes advantage of the GPS information on board and experimental results show that it significantly outperforms OLSR in routing under frequent network topology changes. | Ad hoc |
| Transport layer | Transport control protocol (TCP) | Multipath transmission control protocol [154] | It is a multipath extension of TCP, which has a congestion control algorithm similar to TCP but modified to accommodate multiple paths and the principles of it. | Star; Ad hoc |
| | | Space communications protocol standards-transport protocol [155] | It is an extension and modification of the TCP/IP (i.e., transmission control protocol/internet protocol) for the high bit error rate, long delay, and asymmetrical space environment. | Star; Ad hoc |
| Cross-layer | Cross-layer protocol | Intelligent MAC for UAV (IMAC-UAV) with Directional OLSR (DOLSR) [156] | It uses IMAC-UAV as the MAC layer and DOLSR as the network layer protocol for directed antennas. The first three layers communicate through the shared data set. It reduces end-to-end delay with respect to original OLSR network protocol. | Ad hoc |
| | | Meshed-tree algorithm [157] | It integrates the MAC layer and the network layer in a single protocol, which can form the clusters, route the data from UAVs to the cluster heads and schedule the time slots in a TDMA based MAC layer. | Star; Ad hoc |

Nevertheless, computation coordination should not be confined by the network topology. Under the infrastructure-based or star topology, although the network is dominated by the central entities, the UAV individuals could still make their own decisions in a distributed consensus-based way after obtaining the global information (i.e., from the GCS under the infrastructure-based topology) or the cluster information (i.e., from the central UAVs under the star topology). In [162], the authors provide a distributed path consensus algorithm for multi-robot coordination with consideration of complexities and constraints of the mission scenario. Authors in [163] build an integrated mission planning system consisting of the negotiation-based task allocation and the intersection-based path planning. In [164], a hierarchical path generation scheme for UAVs is proposed to improve the adaptability and performance of the distributed mission planning system.

*3) Cooperative control:* Cooperative control guarantees that each UAV could fly at a proper location to perform complicated missions cooperatively. It reflects and maintains the relationship of the position and velocity between each UAV and its counterparts. For USL, the cooperative control firstly needs to ensure the flying safety of the UAVs in regardless of the application scenarios, i.e., avoid physical collision or collision tendency between them or with the obstacles. Further, sometimes, these UAVs need to keep a regular or specific pattern, i.e., doing the formation control or collective motion, when they fly to a common destination, or for efficiently executing the tasks.

There are lots of researches on achieving the basic collision avoidance for multiple UAVs [165]. They can be classified into two categories: *i)* predefined route and *ii)* real-time planning. In the first category, UAVs are scheduled by a central entity, which considers obstacles, restricted airspace and altitudes limitations [158], [159]. Collisions among UAVs are avoided by using a scheduling algorithm. This method is easy to implement and requires little to the dynamic programming ability of the individual UAV. However, it cannot deal with the pop-up obstacles well and thus has limited application scenarios. In the second category, there are diverse solutions, including the optimization method [166], sense-and-avoid approach [167], force-field method [168] and so on. The optimization method attempts to find an input that minimizes the performance index to avoid obstacles, by adopting PSO to avoid detected static and pop-up obstacles [169], or combining PSO with GA to solve the optimization problems with a nonlinear objective function [170]. These optimization-based approaches are computationally intensive and require heuristic choice of a termination criterion to guarantee a convergence time. The sense-and-avoid approach prevents collisions by changing the travel direction of the aircraft away from the obstacles with the help of multiple sensors [171], which is essentially similar to a pilot's behavior in a manned aircraft. The simplicity of the sense-and-avoid approach results in low computational requirements and short response time. For the force-field, or potential-field approach, virtual fields around the obstacles and UAVs are introduced. And virtual attractive and repulsive forces created by these fields are used to generate collision avoidance maneuvers [11], [172], [173]. However, local minima may exist in the force field, which cannot be easily addressed [174]. These real-time planning methods are good at dealing with emergency cases, and perform better in unknown and evil environment.

Collective motion indicates that a set of individuals moves together as a group in a cohesive way [175]. It can be observed in the nature, such as fish schools, bird flocks and wildebeest herds. Collective motion serves not only to move a group of UAVs from one place to another, but to perform more complex tasks, such as light show, collective mapping and searching. [33] surveys the collective movement of mobile robots, including the classification and characterization of formation types, architecture review and promising applications. The literatures on collective motion can be classified into three categories: *i)* with a leader (e.g., wild goose flocks). There is normally a single leader, and other group members are followers who follow the leader's movement. [176] and [177] try to model the collective motion of the nature creatures with leaders; *ii)* without a leader (e.g., insect swarms and fish schools). All of the members move individually and play an equal role while maintaining contact with each other. The objectives of both [165] and [178] are to model the collective motion without leaders; *iii)* specific applications. It includes the works that try to apply collective motion to some specific scenarios, such as tracking targets [179], [180], forming specific topologies and patterns [172], [181], maximizing the coverage area [182], [183], and maintaining network connections [184], [185].

(c) Applications

Compared with UCL, USL is much more robust and flexible through UAV redundancy, and with broader activity range considering the loose single-hop connection with the GCS. USL can be applied to many fields, including: *i)* network coverage, where multiple interconnected UAVs form an airborne network to offer wireless access for the ground users/sensors [11], [186], [187], statically or dynamically. Or they can construct a line relay to link up two remote areas [2] ; *ii)* sensing and actuation coverage in a large area, where multiple UAVs could cooperatively execute monitoring and tracking mission, such as search and rescue or hitting [5], environment monitoring [188] and border surveillance [189]. The missions could be efficiently performed through the information exchanging among the UAVs.

## C. System of system level

A UAV cloud service platform can be built to absorb diverse USLs or UCLs together, through which the UAV resources are integrated and scheduled efficiently. The cloud service platform facilitates the automatic flow of the multi-source heterogenous data by integrating, exchanging and sharing. Thus, the cross-system and cross-platform interconnection, interworking and interoperability among the UAVs can be achieved. Also, a closed loop of the comprehensive perception, in-depth analysis, decision making and precise actuation is guaranteed.

In USoS, each UAV will access the cloud service platform. This platform maintains the UAVs' information about their resources and services, and also provides a uniform interface to invoke them. One one hand, the UAVs contribute themselves to the platform, e.g., the sensed information, computing resources and actuation abilities. On the other hand, they could acquire services from it, e.g., the firmware upgrade, remote main-tenance and service subscription. Furthermore, the platform would also provide the UAVs as services to the users, e.g., the sensing, actuation, camera capturing and video recording services. Thus, USoS is much like a UAV ecosystem where all kinds of UAVs flourish and are  orchestrated.

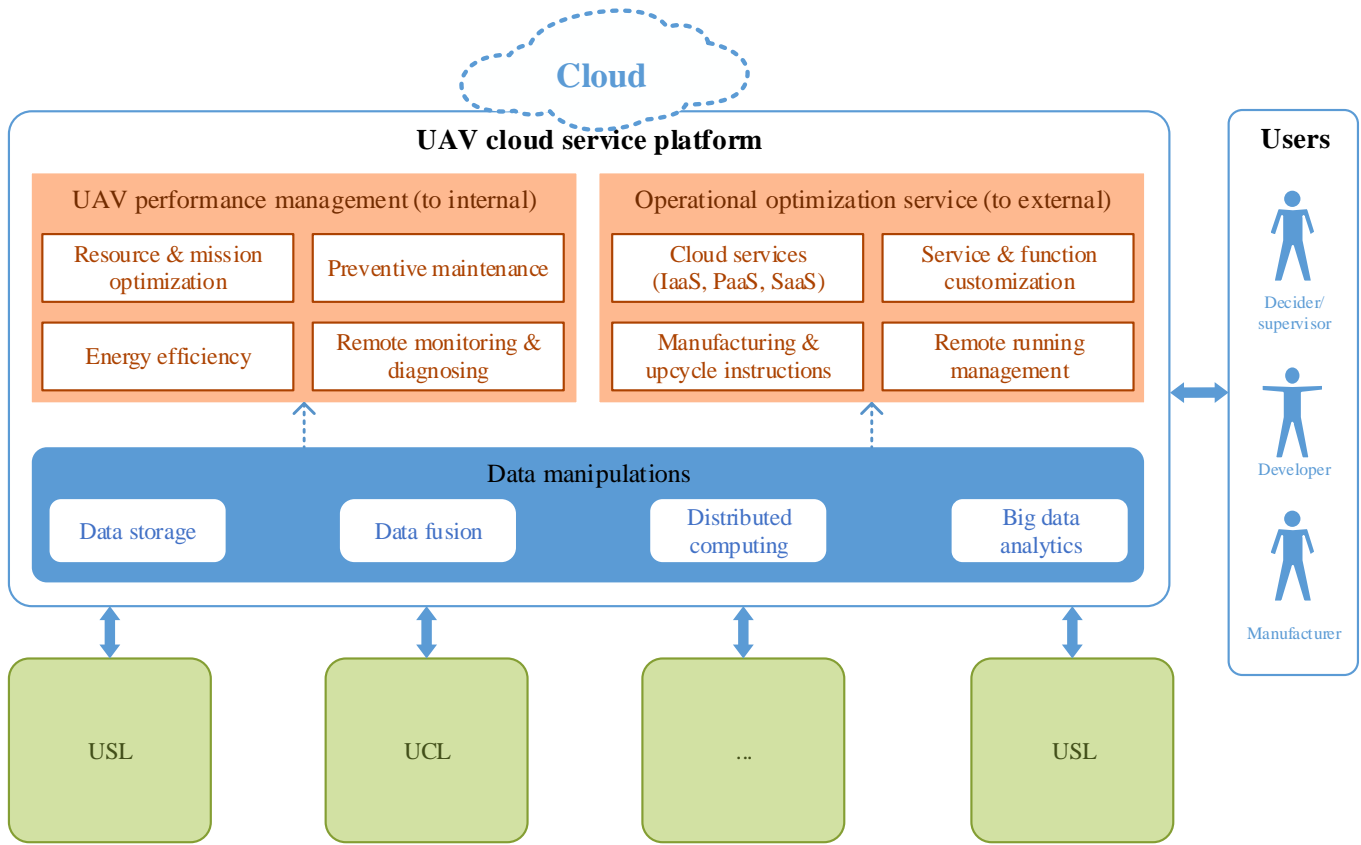

Fig. 14.  The constitute architecture of the  USoS.

**(a) Architecture**

*1) Constitute architecture:* The constitute architecture of the USoS is shown in Fig. 14. USoS emphasizes much on  the data and service convergence, to optimize the UAVs' operation performance internally and provide services to the external. The system consists of three abstraction layers [190], namely:

- **UAV layer**: This layer provides system resources for the users, in other words, UAVs as a service [191]. The UAV layer focuses on the hardware interaction by using the middleware and data links. The data links, which may be built over various transport protocols, allow the exchange of predefined messages between the UAVs and the GCS and among the UAVs. Together, the middleware and the data link provide high-level interfaces for the users to monitor and control UAVs without the need of direct programming;
- **Cloud service layer**: This layer focuses on realizing cloud services using three sets of components, i.e., communi-cation interfaces, data manipulations (e.g., data storages, data fusion, distributed computing and big data analytics) and high-level applications. In detail, *i)* communication interfaces include the network interfaces and web ser-vices. The former is used for handling continuous stream-s, and the latter is used for sending control commands to UAVs and receiving information from the cloud; *ii)* the data streams originating from UAVs are partially conveyed to and stored in the  cloud  service platform, and others are distributed among  each  individual.  So, big data analytics and distributed computing can be adopted to provide support for the data services and advanced applications; *iii)* high-level applications include the UAV performance management and operational op-timization service. The former mainly includes resource and mission optimization, preventive maintenance, energy efficiency, and remote monitoring and diagnosing. The latter emerges in the form of the cloud services, service and function customization, manufacturing and upcycle instructions, and remote running management. Through this layer, each individual's state could be manipulated re-spectively and optimized collaboratively, so as to achieve the optimal resource  configuration;
- **User layer**: This layer provides  interfaces  for  all  the end users, including the deciders/supervisors,  application developers and manufacturers. In detail, *i)* for the decider-s/supervisors, this layer executes client/browser-side web applications that provide an interface to the cloud services and UAV layers. The applications allow them to  register, remotely monitor and control UAVs and missions, by defining and modifying mission parameters provided by the cloud; *ii)* for the developers, the layer provides several application programming interfaces (APIs) of different programming languages for straightforward development of drone applications. These new software and algorithms can be loaded to the UAVs through firmware and software upgrade online; *iii)* for the manufacturers, they could hear the failures of each UAV and give the maintenance measures in good time through the interfaces. Besides, a comprehensive report created through the big data analytics would provide advices with respect to the firmware upgrades, operating maintenance, performance metrics and subsequent upcycle of the UAVs.

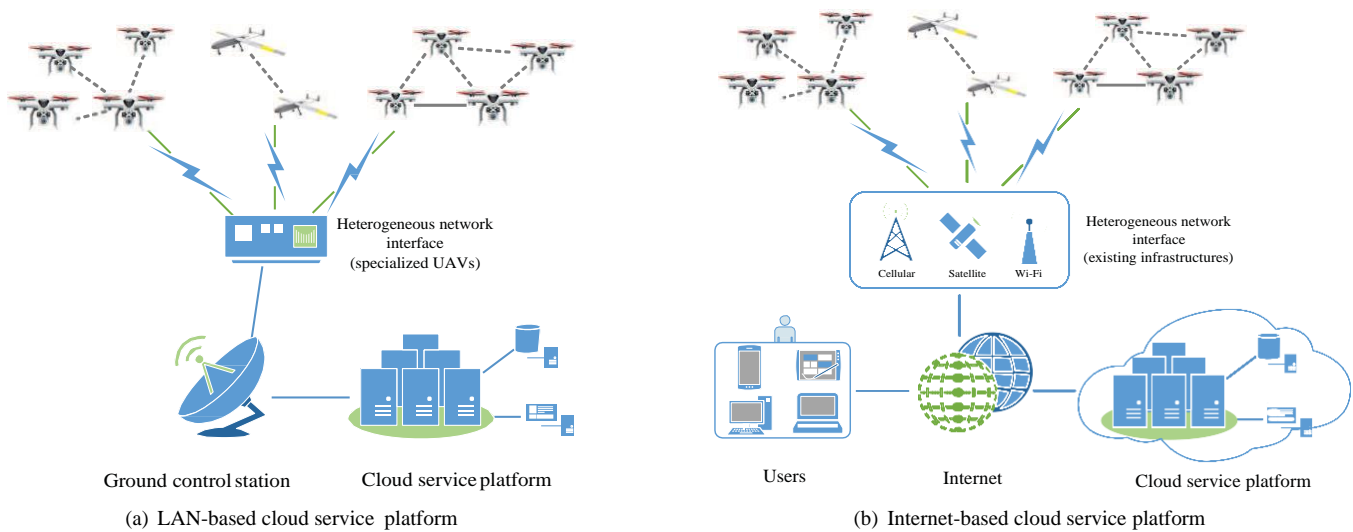

(a) LAN-based cloud service platform  (b) Internet-based cloud service platform

Fig. 15. Two representative organization architectures of the USoS.

*2) Organization architecture:* Compared with UCL and USL, the organization architecture of USoS is much more complicated. Based on the networking among the UAVs in USL, the UAVs need to be connected to a cloud service platform further. So, how USoS is organized is much determined by where the cloud service platform is located. Here, we consider two scenarios where the cloud service platform is built based on the local area network (LAN) or the Internet respectively. The differences are as follows: i) for the former, the UAVs can be directly connected to the platform through the local area network. However, for the latter, all UAVs first need to access the Internet, which may resort to the existing infrastructures, such as the cellular base stations and communication satellites; ii) the former is applicable to the military missions and the platform could be directly deployed on the GCS. However, the latter is more popular in civilian areas since the platform could be deployed in a commercial cloud platform where all kinds of UAVs, developers and amateurs could participate anytime and anywhere.

Fig. 15. describes two representative organization architectures of USoS, with a LAN-based platform (Fig. 15 (a)) or an Internet-based platform (Fig. 15 (b)). In both scenarios, the heterogeneous network interfaces are indispensable, which are responsible for establishing the convergence between the heterogeneous UAV networks and the cloud service platform. For the former, the heterogeneous network interfaces can be deployed on the specialized UAVs, and for the latter, the interfaces are usually developed on the existing infrastructures (e.g., Cellular, satellites and Wi-Fi).

(b) Key techniques

The perceived data in USoS is much richer compared with the other two hierarchies. The information and knowledge extracted from the multi-dimensional data will enable greater insight and capabilities of decision making and resource optimization for the UAVs and the deciders. Thus, some new data processing techniques are needed to extract the potential value of the huge data. And through data services, the platform and the UAVs' abilities to control and optimize the resources can be improved. To sum up, the accessorial technical requirements of the USoS, based on those in USL, can be summarized as: *i)* flexible distributed computation; *ii)* resource and service provision (to the external).

*1) Flexible distributed computation:* The emerging computation technologies, including the edge computing, fog computing and cloud computing, would help to build the cloud service platform by flexibly processing the data according to the mission requirements. They enable the platform to provide data services, and further, the customized and specialized intelligent services, to the external. Fig. 16 describes the computation technologies required in USoS.

- *Cloud computing*: Cloud computing provides users with a wide range of virtually unlimited resources, including the infrastructures, platform and software services. All data from physical assets is conveyed to the cloud for storage and advanced analysis. Since cloud has considerable computing power compared to the devices at the network edge, shifting computation-intensive tasks to the cloud computing platform is an effective approach for data processing. There are many researches focusing on developing the cloud-based system to control, manage, or communicate with drones over the Internet, such as Dronemap Planner [192], Robot Cloud [193], Cloudroid [194], Context Aware Cloud Robotics System [195], and so on. Cloud computing will boost the big data analytics due to its advantages in terms of the computation power and the amount of the acquired data, which can bring profound changes to the global UAV industry. Indeed, Cloud computing has been an outstanding service provider which could serve delay-tolerant applications with ease. Nevertheless, delivering of massive data loads over a network puts enormous stress on the network resources. Besides, it is not friendly to support the requirements of location awareness, low latency and mobility, since in some cases, it is essential to process data near its source. For example, in the safety-critical applications, including battles, healthcare and so on, a delay of even milliseconds can be fatal.

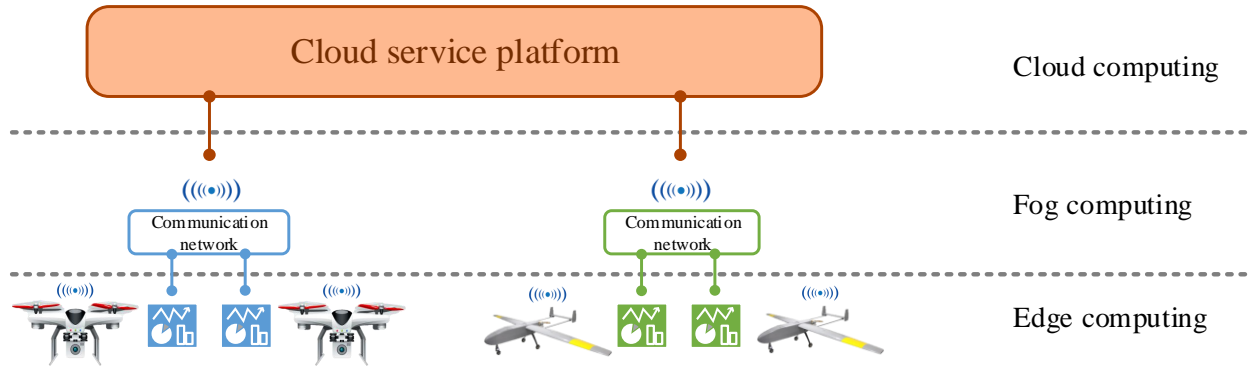

Fig. 16. The computation technologies needed in the USoS.

- *Fog computing*: USoS is a complex and vast control system. So, a local UAV network (e.g., a USL) also needs to perform collaborative calculation for each UAV to cooperatively control them. As a new generation of distributed computing technique, fog computing converges the data, data processing and application programs in the devices located on the edge of the network [196]. These devices may include the intelligent routers (they can be the specific UAVs) and local GCS. The data storage and processing rely much on the local devices other than the remote servers. Fog computing extends cloud computing to the end devices, to better support time and location-critical, massive scale and latency sensitive applications. Accordingly, the network traffic and computation load of the data centers can be effectively reduced. Fog computing can be adopted inside the UCL or USL to cope with the large amount of data generated by the local network, for example, preprocessing the data to reduce its size and enhance its value. In [197], a UAV-based fog computing platform is designed, where several UAVs are deployed as fog computing nodes to provide more localized real-time monitoring, control and optimization for the internet of things applications.

- *Edge computing*: Edge computing is an open computing paradigm, which supports data calculation at the edge

of the network (i.e., the UAVs), in proximity of the objectives or data sources. Computation in proximity eliminates many drawbacks resulting from the computation in the remote cloud. It tremendously reduces the data flow, bandwidth utilization and latency by processing the larger chunks of data at the edge, rather than redirecting them to the cloud [198]. Therefore, it could meet key requirements of the UAV applications with respect to agile connections, real-time data analysis, and security and privacy protection. In addition, UAVs, the leading role of the edge computing, are the data acquisition units for valuable data which is required in the cloud, and also the final execution units of the services which are provides by the could. Thus, edge computing will better support the intelligent cloud services. In [199], the authors have adopted the edge and fog computing principles to the UAV-based forest fire detection application through a hierarchical architecture.

This three-layer computation ecosystem combines the powerful computation abilities of the cloud computing, the rich resources of the fog computing, and the sensing and actuation capabilities of the edge computing (i.e., UAVs). Shifting lots of computation tasks from the remote cloud to the network's edge is expected to decrease the latency and load of the network. However, fog computing and edge computing are not going to substitute the traditional clouds, since that the amount of computation power and storage they could offer is much less than those of traditional clouds. Instead, the aim is to complement the traditional cloud data centers by running some delay-sensitive applications at the edge of the network [200]. All the three computation technologies will contribute to the cloud service platform and the USoS. And they should be tuned well when allocating computing tasks, i.e., putting the data and tasks to the appropriate computation entities, far to the remote cloud or on the near edge devices.

*2) Resource and service provision:* By integrating all the available UAVs, the cloud service platform could ubiquitously provide resources and services to the external requesters through the APIs. UAVs can be mapped to the three cloud service models to combine UAV resources with other cloud features [201]. The framework of the UAV cloud service is shown in Fig. 17.

- **UAV IaaS**: The infrastructure as a service (IaaS) model includes UAV hardware components and some other components. The UAV hardware components consist of their sensors, actuators, payloads, internal memory, processors and other internal resources (which are represented by R1-Rn in Fig. 17). Other components may include any external entities which would provide resources or services, such as the ground sensors and objects connected to the cloud, or the cloud computing infrastructures (e.g., the high-performance storage, servers and processors). These are all managed through APIs to the Platform as a Service (PaaS).

- **UAV PaaS**: The PaaS is modelled as the middleware to isolate the infrastructure layer from the application layer. It offers resources as services to the application layer. The platform includes UAV resources and services as well as cloud services through APIs. UAV resources and services offer specific data from sensors, or perform some actions using certain actuators, for example, acquiring a current reading from the temperature and gas sensors, performing pesticide spraying, or capturing images and videos. Besides, database management system (DBMS) is needed for information storage.

- **UAV SaaS**: The software as a service (SaaS) is a lightweight software application available online. It is built on top of the PaaS through standard APIs. The applications are developed for users to request certain UAV missions. For example, UAVs could be requested to spray crops for a specific agricultural area through the software. The users only need to access the application to specify the location and size of the area, and then ask for crop spraying from the UAVs. The application may also offer monitoring interfaces for the users to track the progress of the mission.

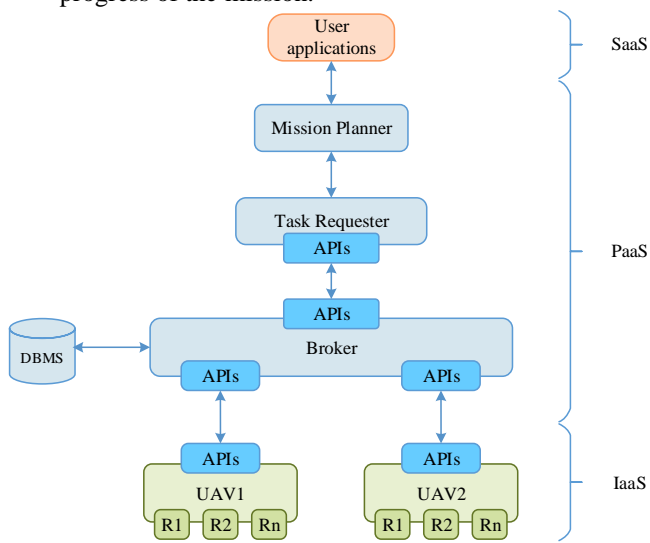

Fig. 17. UAV cloud service framework.

Developing the PaaS layer, i.e., the middleware, is of great importance, which integrates UAVs to the cloud and provides an efficient layer to build applications on it (please to Fig. 17).

All the components, including the UAV themselves and some other computing components, could be developed as services and integrated into the applications if necessary. These services include UAV services and collaborative services. The former is offered and used based on the UAVs' capabilities, while the latter is required for any type of collaborative UAVs. Building applications upon these services reduces the time and cost of developing collaborative UAV-related applications.

- **UAV services**: They are accessed according to the available resources in UAVs and the tasks required for the missions. UAVs may employ one or more of them. They may include: *i) sensing service*, which acquires data from the sensors and conveys it to the broker service. The request of this service could involve either acquiring the value of the sensors, or setting a threshold to be triggered when the sensors meets the condition; *ii) actuation service*, which relies on the output devices on the UAVs, such as lights and liquid/gas valves (e.g., in spraying missions). A set of actuation services can be provided in each UAV; *iii) energy monitoring service*, which is necessary since many decisions and task allocations are made based upon the UAV's energy level; *iv) location monitoring service*, which is needed because of that UAVs' locations play an important role in allocating tasks. For example, if a UAV is currently near the mission location, it is preferable to be chosen; *v) status service*, which would monitor the status of UAVs and return information about the resources.

- **Collaborative services**: They manage the distribution of UAVs to accomplish a mission, and may include: *i) broker service*, which maintains all UAV resources and services in a database. When a request is given, the broker service will search for UAVs with those resources or services. Then, it queries the UAVs to find the suitable ones and allocates tasks to them with the requested parameters, based on the resources, locations, energy levels and other considerations; *ii) task requester service*, which is responsible for requesting these tasks from the broker service. The task requester does not have knowledge about the UAVs and their capabilities. However, it could still request a certain resource and specify its parameters according to the schedule; *iii) mission planner service*, which focuses on decomposing the mission into tasks, and subsequently defining the functions, parameters and resources required to perform the mission according to the current and expected conditions;

The UAV cloud service platform architecture consists of both the UAV services and the collaborative services that can be accessed through Web service APIs. There are various types of Web services in different architectures, therefore, the platform should follow the specified requirements and considerations.

(c) Applications

Considering the ubiquitousness characteristic of the cloud service platform, UAV networks of any hierarchy could access it from any where, at any time and through any method. They integrate themselves into it as well as enjoy and flourish from it. In military areas, the USoS could be deployed for a joint operation which involves all kind of heterogenous UAVs and other unmanned/unmanned vehicles. In civilian areas, the cloud service platform could be built based on or extended from the off-the-shelf Internet-based cloud platforms, such as the Dronemap Planner. So, the USoS could be used to various applications so long as the UAVs are registered into the cloud platform with the Internet access. Maybe, someday, the shared drone business will prosper with the development of the UAV cloud. These applications can be cross-industry, including the agriculture, transportation, meteorology, geography and so on.

TABLE X
COMPARISON OF THE UAV NETWORKS UNDER THE THREE HIERARCHIES

| Hierarchy | Cell level | System level | System of system level |
|---|---|---|---|
| Constitutes | Hardware + Software | Hardware + Software + Communication network | Hardware + Software + Communication network + Platform |
| Descriptions | Single UAV or multiple noninteractive UAVs are connected to the GCS. | Multiple interactive UAVs are connected to the GCS through one or several UAVs | Multiple UAV networks of system level or cell level are connected to a cloud service platform through the GCS or the existing infrastructures, such as the cellular base stations and Wi-Fi |
| Topologies | Infrastructure-based | Infrastructure-based; Star; Ad hoc | Mesh (heterogeneous network) |
| Communication entities | Between UAV and the GCS | Between UAVs and UAVs, UAVs and the GCS | Between UAVs and UAVs (in a USL), UAVs and the GCS, UAVs and other existing infrastructures |
| Link types | Point-to-point/point-to-multipoint | Point-to-point; End-to-end | Point-to-point; End-to-end |
| Channels | UAV-ground | UAV-ground; UAV-UAV | UAV-ground; UAV-UAV |
| Computation entities/modes | On the UAV locally, or on the remote GCS | Distributed/centralized on the UAVs locally, or centralized on the remote GCS | Distributed/centralized locally on the UAVs (edge and fog), or centralized on the cloud platform |
| Control | Individual flight control | Formation control/collective motion | Individual and formation control |
| Closed loop | Small | Medium | Large |
| Advantages | Agility; Low cost; Simple communication and control | Reliability and survivability through redundancy; Large mission area | Ubiquitous access; Data, information, knowledge and experience accumulation |
| Disadvantages | Vulnerable; Limited mission area | Complex communication networking and coordinated control | Complex heterogeneous network; Internet-dependent |
| Typical applications | Package dispatching; Agricultural plant protection; Aerial photography; UAV-aided relaying | Aerial base stations; Search and rescue; Large-scale surveillance and reconnaissance/tracking/hitting | Shared drone in various fields; Joint military operations |

In Table X, we comprehensively compare the UAV networks under the three hierarchies.

## V. COUPLING EFFECTS IN THE UAV NETWORKS

In the UAV networks, with a dataflow-based closed loop being built, the cyber-domain components and the physical-domain components are tightly coupled, as shown in Fig. 18. The term "coupling effect" here has two meanings. First, it means the macroscopic data flows between the cyber domain and the physical domain. The data flows are introduced into the cyber domain by the sensors from the physical world, and are finally fed to the actuators and take an effect on the physical world. Second, it implies the mutual influences and dependencies among each of the components in the cyber/physical domain on a micro level. Understanding these coupling effects can greatly help us to properly tune the crucial yet usually restricted network resources with respect to the sensing, communication, computation and control [202].

### A. Computation and communication

Communication and computation will be promoted by each other. Fig. 19 describes the relations between the communication and the computation.

*1) Communication contributes to computation:* Intuitively, communication contributes to computation through the data flow. The input data for computing or decision-making is conveyed by the communication network (e.g., the state dissemination among the individuals for collision avoidance, path planning and task allocation). And the computation outputs of each entity, i.e., the decisions, are usually shared to others through the communication network for reaching a consensus.

Sometimes, the performance of the communication may confine the computation ability, especially for the tasks that require real-time and high-throughput inputs (e.g., the collision avoidance under high speed). So, the performance improvement of the communication (e.g., the throughput increase and latency reduction) will give a positive feedback to the computation ability, and may eliminate the bottleneck of the computing speed brought by the communication. The communication also makes global optimization much easier. For example, recently, the multi-agent partially observable Markov decision process (POMDP) has been proposed as a mechanism for coordinating teams of agents (actually, the UAVs can be regarded as agents since they acquire all the features of the agents). It is known to be NEXP-complete [203], which makes optimal policy-generation intractable. However, communication provides a valuable tool for improving the tractability of making the optimal policies, by building a consensus belief among the individuals. Moreover, in a large-scale network consisting of UAVs with high computation power, the cooperative computing and the swarm intelligence could be boosted by the communication network among the individuals.

*2) Computation boosts communication:* Computation will enhance the communication, and is expected to unlock, at least approach, the communication bound, i.e., the Shannon's law. Here, we take the coupling effects between the communication and the physical world as an example, as shown in Fig. 18. The geographical terrain (including the plants, rivers, mountains and buildings) determines the channel model between the UAV transmitter and receiver [204], which further has influence on the communication performance between them. So, what computation needs to do is, first, measure and model the channel (i.e., obtaining the characteristics of the channel) based on its understanding about the circumstances; then, classify the channel model according to the predefined rules (by using some machine learning algorithms); and finally, match and load the exact optimal waveform in the off-the-shelf waveform library if there exists (e.g., the waveform library could be initially built based on serval typical communication environments and further be extended with the optimal waveform decisions made under occasional and unknown environ-ments), or directly make a new waveform decision according to the fresh measurements. Thus, computation facilitates and unlocks the communication ability. For the UAVs with learning abilities, the more this process iterates, the better the nodes will communicate in unknown environments.

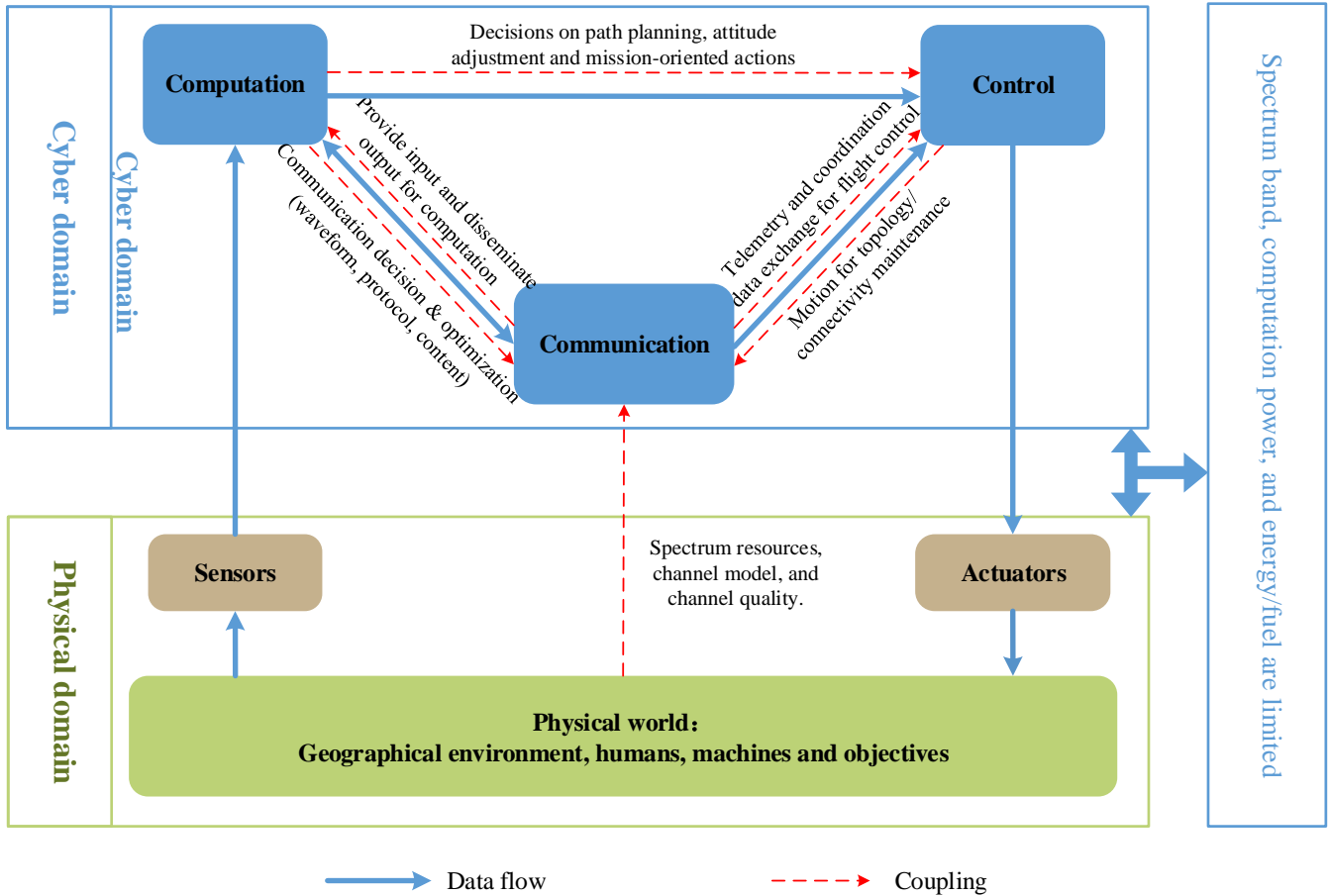

Fig. 18. The coupling effects in the cyber and the physical domains.

There are many researches focusing on embedding the computation into communication to improve the communication and the networking performance for the UAVs, in other words, exchanging for communication using the computation. They can be classified into two categories, which adopt quite different methods. The first category tries to enhance the performance of the exact communication and networking themselves, from different layers of the network. Some intelli-gent algorithms are adopted to cope with the challenges, such as low latency and high transmission rate requirements, which are brought by the intrinsic features of the UAVs (e.g., the frequent topology changes and status interactions, and also the high-resolution image or video transmissions). In the second category, the works concentrate on lessening the communication quantity and overhead in the whole system, considering the cost of communication. The agent decision technology is introduced to let UAVs decide whether to [205], [206], when to [207], what to [208], and whom to [209] communicate. This category makes the communication more precise, and thus the most needed information being transmitted to the most wanted nodes at the most appropriate time can be achieved. In turn, communication resources can be saved and the performance of the necessary communications can be improved.

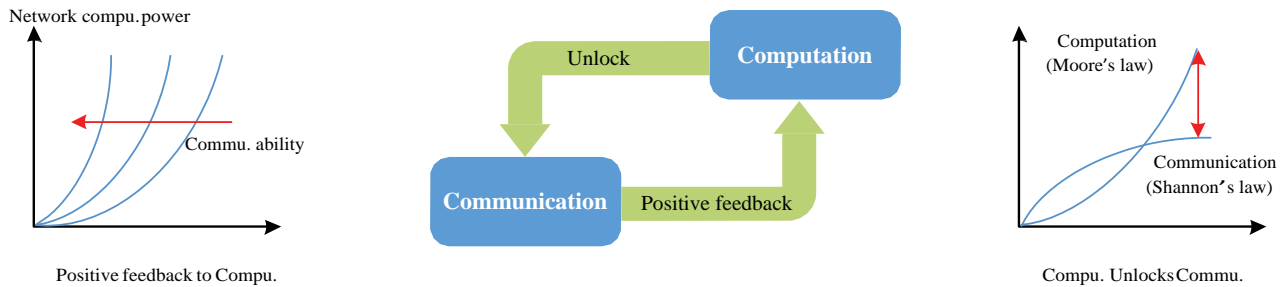

Fig. 19. The promotion relations between the communication and the computation.

- **Intelligent algorithms boost the communication and networking**: The intelligent algorithms could be exploited to improve the performance of the communication and networking from different layers of the network. For example, *i)* in the physical layer, the machine learning algorithms are usually used in channel modelling [210], channel estimation [211] signal detection [212] and so on; *ii)* in the data link layer, the Markov decision process (MDP) is usually formulated to choose the best medium access strategy for overcoming the channel congestion [213], to maximize the network throughput [214], or to guarantee the delay and Qos enhanced data transmission [215]. For the frame-based ALOHA scheme, reinforcement learning is implemented as an intelligent slot assignment strategy to avoid collisions with minimal additional overheads [216]; *iii)* in the network layer, the reinforcement learning can cope with the dynamic conditions in the UAV networks. A survey of learning-based routing algorithms for the ad hoc networks can be found in [217], which is enlightening to the routing in the UAV networks. Q-routing [218], the first routing algorithm based on the reinforcement learning, is proposed to realize the intelligent decision-making and self-optimizing for the packet routing under the changing traffic and topology. The authors in [219] propose a memory-based Q-learning algorithm called predictive Q-routing (PQ-routing) for adaptive traffic control. They attempt to address two problems encountered in Q-routing, namely, the inability to fine-tune routing policies under low network load and the inability to learn new optimal policies under decreasing load conditions. Simulation results show that, PQ-routing is superior to Q-routing in terms of both learning speed and adaptability. There also many other extensions of Q-routing, e.g., full echo Q-routing [220], dual reinforcement Q-routing [221], ant-based Q-Routing [222], gradient ascent Q-routing [223], Q-probabilistic routing [224] and simulated annealing based hierarchical Q-routing [225]. In [142], a self-learning routing protocol based on reinforcement learning (RLSRP), specific for the FANET, is studied. In the protocol, all flying nodes will exchange their status information with the FANET to update the stored data, and make decisions on the routing path that has the shortest delivery delay.

- **Multi-agent decisions reduce the communication overhead**: Generally, communication is a limited resource and therefore cannot be treated as being free. Sometimes, communication is unavailable at times, or it may be dangerous to communicate. Therefore, the decisions in the multi-agent system should be made with bounded communication [226], without sacrificing the task performance. To use communication effectively, multi-agent teams should be able to calculate a value of communicating, and then trade off the benefit that can be achieved through communication with the cost of communicating [209]. [227] presents a novel model of rational communication, which uses reward shaping to value communications, and employ this valuation in the decentralized POMDP policy generation. The authors in [207] present an algorithm for making execution-time decisions about when to communicate. The algorithm inserts communication actions into the team execution only when it will improve the team performance, and thus reduces the communication amount needed for successful execution of a decentralized policy. Further, they propose an algorithm that builds their previous works to address the problem of what to communicate in [208], by answering which state features are most relevant to the team performance and which observations should be transmitted. In [205], the authors have developed an online, decentralized communication policy, i.e., ConTaCT. The policy enables agents to decide whether or not to communicate during the time-critical collaborative tasks in unknown, deterministic environments. And it can substantially reduce the communications among agents while achieving comparable task performance compared to other multi-agent communication policies.

## B. Computation and control

In the UAV networks, control highly depends on computation since that the control decision making are usually undertaken by the onboard computer, and the computation outputs, i.e., the digital control inputs, are directly translated into continuous signals and fed to the actuators. In this part, we focus on the couplings between the onboard computation power and the flight control of the UAVs.

The promotion of the computation power will facilitate the flight control of the UAVs, by enabling real-time optimal path planning and formation control, making faster response to the external disturbances, or decreasing the system model dependency by adopting some learning algorithms [228], [229]. Of course, the computation power is determined by both the hardware and software components. On one hand, the flight control modules with high-performance processors can be equipped to boost the computation ability of individual UAVs, such as the Qualcomm Snapdragon Flight, Nvidia Jetson TX1, Leadcore LC1860 and so on (as illustrated in Part B, Section III). On the other hand, one possible approach that would relax the difficulties in the nonlinear control design is to adopt some learning algorithms, which allows the training of suitable control actions. So, the intelligent algorithms could also be developed to deal with the flight control and formation control problems more efficiently, which, in reverse, exactly boosts the utilization of the high-performance processors on the UAVs.

*1) Computation-enhanced flight control:* The intelligent flight control methods can overcome the deficiency of the classic controllers by combining the idea of learning with the conventional control algorithms together [29]. The main characteristic of these methods is that no prior mathematical knowledge of the model is required to design a controller, but some of the flight data are necessary to train the system [28]. The intelligent flight control methods open up novel control theories. Thus, they may flourish with the development of the artificial intelligent algorithms. Here, we summarize the learning-based intelligent flight control algorithms as follows.

- *Fuzzy logic-based flight control algorithms*: Fuzzy logic considers the qualitative knowledge of the designers and/or the human pilots, and translates it into fuzzy rules in the control systems. The main advantages of fuzzy control stem from its model without accurate accused process. So, such control method has strong robustness, good adaptability, fault tolerance and is easy to realize. But the shortcoming is that there may be some steady-state error about the actual control effect due to lack of integral link. In [89], the authors have designed and implemented intelligent fuzzy logic controllers for the quadrotors. Other researchers have combined the intelligent fuzzy with the traditional controllers. For example, in [230], the authors make use of the fuzzy PID controller to reduce the whole flight control system's influence from the external disturbance, and improve the robustness of the system. In [231], a fuzzy PID controller is designed, which achieves automatically path tracking and reduces the influence of external disturbances.

- *Artificial neural network-based flight control algorithms*: Artificial neural network (ANN) is biologically inspired by the central nervous system. It will boost the classical controllers, e.g., PID, by adjusting the parameters online to deal with the unknown variable payload issues [232]. It has good self-learning ability, strong robustness and high searching speed. However, the learning phase tends to be long and it is difficult to explain their reasoning process. So, both the offline trained and the online learned can be combined to ensure the fast learning and robustness [233]. The ANN output feedback control law [228] has been implemented on a quadrotor for tracking a desired trajectory. It could learn the complete dynamics of the quadrotors online, including uncertain nonlinear terms like the aerodynamic friction and blade flapping. In [233], two back propagation networks are used in the control system to achieve the model identification and control.

- *Reinforcement learning-based flight control algorithms*: Reinforcement learning is to learn what to do, i.e., how to map states to policies, so as to maximize a numerical reward. The main advantages of this method are model free and online learning, which make it robust to the uncertainties and external noises. Nevertheless, the environment may be partially observed, and the UAVs may get incomplete perception to the environment. In recent years, there are some works using reinforcement learning for the quadrotors [229], [234], [235]. For example, in [234], the reinforcement learning control is presented for the outdoor altitude control of the multi-agent quadrotor testbed, which brings a significant improvement over the classical linear control techniques in dealing with the nonlinear disturbances.

*2) Computation-enhanced formation control:* The formation control deals with the cooperative flight within a large-scale UAV swarm (also referred to as a multi-agent system). It can be enhanced by adopting some intelligent algorithms, including the bio-inspired algorithms and the artificial intelligent algorithms. The former derives from the animal behaviors in a cooperative team, which reflects the characteristics of the UAV swarm very well, i.e., heading for a goal while keeping a safety distance, a regular layout or formation reconfiguration [236] according to the information acquired. The latter mimics the thinking methods of the human, which could efficiently solve the problems with lower cost compared to the traditional optimization algorithms.

- *Bio-inspired algorithms-based formation control*: As one of the most popular bio-inspired algorithms, PSO imitates birds or fishes group foraging behaviors, which exactly reflects the 3D flying nature of the UAVs from the motion perspective. So, there are many papers applying PSO to the complicated formation optimization problems in multiple UAVs, such as [237]–[239]. For example, in [238], a hybrid PSO algorithm which consists of the continuous and discrete PSO algorithms is introduced to solve the trajectory planning problem of multi-robot formation. Besides, there are some researches using GA to optimize the parameters of the multi-UAV formation control method based on artificial physics [240], to make motion plan for the artificial potential-based formation control [241], or to combine with the particle filter so as to solve the dynamic optimal formation control problem [242]. Of course, the GA could cooperate with the PSO to solve the complicated formation reconfiguration problem in 3D space with strict constraints and mutual interferences. In [170], the authors model the multi-UAV formation reconfiguration as a parameter optimization problem, and propose a hybrid algorithm of PSO and GA to solve it. This new approach combines the traits of PSO and GA, and can find the time-optimal solutions simultaneously. Simulation results show that the hybrid algorithm outperforms the basic PSO under complicated environments.

- *Artificial intelligent algorithms-based formation control*: Usually, the optimal formation control is achieved by solving the Hamilton-Jacobi-Bellman (HJB) equation. However, it is very hard because of the unknown dynamic and inherent nonlinearity. When it comes to the multi-agent systems, it will become more complicated owing to the state coupling problems in the control design [243]. Artificial intelligent algorithms provide a new method to solve the formation control problem well, especially for the scenarios where some system models are unknown, or hard to build theoretically. For example, the neural network could learn the complete dynamics of the UAVs, including the unmodeled dynamics (e.g., the aerodynamic friction) [244]. So, a neural network-based cooperative controller could be designed to achieve and maintain the desired formation shape in the presence of unmodeled dynamics and bounded unknown disturbances [245]. In [246], a neural network-based adaptive consensus control is designed for a team of fixed-wing UAVs. And in [247], a new neural network-based control algorithm which uses graph rigidity and relative positions of the vehicles is proposed to address the decentralized formation control problem of the unmanned vehicles in 3D space. However, the neural network needs to be trained previously with enormous teaching samples before applying it, which might be costly. Reinforcement learning only has a reinforcement signal, which is used for evaluating the behavior and further strengthening the tendency of the behaviors that could lead to positive reward. Thus, it is much preferred when solving the formation control problem in a distributed way. In [248], the authors have presented a novel multi-objective reinforcement learning formulation of the decentralized formation control problem for the fixed-wing UAV swarms. Both in [249] and [243], a method of applying the reinforcement learning algorithm to a leader-follower formation control scenario is proposed. As for the fuzzy logic controller, reinforcement learning could be combined to improve the learning speed of the formation behavior, by adjusting the weighting factors of the behavior fusion [250], or estimating the unknown nonlinear dynamics [243].

### C. Communication and control

Communication and control are also two tightly coupled components, and they will constrain and promote each other. As the intrinsic feature of the UAVs, the three-dimension mobility is dominated by the flight control, which brings more challenges to the communication and networking. In return, the flight control also creates a novel method to solve the problems in communication by integrating the physical domain, e.g., location changes, other than only adjusting the communication parameters. As two main power consumers, communication and control should be planned and tuned well considering the energy constraint of the UAVs [251].

#### 1) The dependency between communication and control:
The flight control of each UAV or the formation control of the whole swarm depends on the status interaction among the individuals (e.g., the telemetry data), which is indispensable for making flight decisions [136]. It has been shown that a communication scheme needs to be adopted to increase the aggregated maneuverability of mobile agents [252]. Even for the UAVs with high autonomy, the flight command transmitted from the GCS is also necessary in emergency. At the same time, the performance of the communication will have significant effect on the flight control of the UAVs. For example, the communication delay may bring the risk of collisions in a UAV swarm [253].

In return, the communication also rely on the flight control. The communication network among the UAVs is guaranteed by the optimal topology control, or at least, the proper distance maintenance between each node, which should be considered when doing the formation control. A large distance between two nodes may bring more transmission power or number of hops. This can be alleviated by shortening the distance between them, i.e., one fly to approach the other (without violating with the mission), which reduces the transmission power while harvesting a comparable communication performance. However, both enhancing the transmission power and moving the node will bring extra energy consumption while improving the communication performance. Thus, there may be an optimization tradeoff between the communication and the control under the constrained power. That is how to tune and plan both of them, i.e., how to allocate the limited power to them jointly, to achieve maximum communication improvement while consuming minimum extra power. In fact, only increasing the transmission power may not be reasonable since that higher transmission power means stronger interference to others and lower spectrum reuse. And sometimes, it can be totally unavailing especially when there exist strong shadow effects. All in all, the mobility of the UAVs can be fully utilized to improve the communication.

#### 2) The constraints between communication and control:
In most cases, communication and control are usually planned and optimized under the constraint of the other. And here, we call them communication-constrained control and control-constrained communication respectively, considering *i)* communication introduces delays to the system, which may decrease the performance of flight control algorithms and increase the collision risk among UAVs. Thus, the safe and efficient flight control, including the collision avoidance and the path planning, is usually scheduled under the limited communication, in other words, the limited or delayed status exchange among UAVs; *ii)* the flight control is mainly dominated by the mission and does not always serve for the communication. Thus, the communication should be planned and optimized to keep pace with the flight control, and to solve the consequent troubles, including the fluid topology, intermittent links and Doppler effect. For example, when a UAV swarm is tracking and hitting the randomly moving targets with high speed, they may perform frequent formation transformation for efficiently completing the task, although it is not friendly to the communication.

The existing researches, which aim at dealing with the communication challenges brought by the mobility from different network layers, can be regarded as the solutions to the control-constrained communication problem. And we have surveyed it in Part B, Section IV. As for the communication-constrained control problem, there are many papers concentrating on dealing with it, and they can be classified into two categorized in general:

- **Formation control under limited communication**: The communication constraints are considered, including the delays, limited bandwidth and range [136], [254], [255] and so on. For example, in [254], the authors integrate the leader-follower strategy and virtual leader strategy into an optimal control framework to deal with the communication constraints, in a known and realistic obstacle-laden environment. In [136], assuming limited communication bandwidth and range, the authors have achieved the coordination of UAV motion by implementing a simple behavioral flocking algorithm, which utilizes a tree topology for distributed flight coordination. In [255], the formation control problem of a multi-rotor UAV team with communication delays is addressed. Three control schemes are presented which provide delay-dependent and delay-independent results with constant and time-varying communication delays.

- *Formation control with communication guarantee*: It means that the communication topology and connectivity among the flying UAVs must be ensured in the formation control. Connectivity-preserving formation control of UAVs/multi-agents can be developed in the form of centralized [256], [257] and decentralized [184], [258]–[264]. The latter can be further divided into two main categories. The first is the conservative connectivity preservation, which aims at preserving all the existing links as system topology evolves. Related works include [258]–[261]. The second is the flexible connectivity preservation, which allows the underlying communication topology to switch among different connected topologies. Some links may be removed or added as long as the overall graph is connected. Related works contain [262]–[264]. In [184], a novel decentralized formation control strategy with connectivity preservation is presented, which extends and generalizes the connectivity preservation methods in a flexible way.

*3) The promotions between communication and control:* Communication and control also promote and benefit from each other. Intuitively, the performance improvement of the communication will motivate the formation control by efficiently delivering the status among the UAVs. Thus, dramatically, flight control of the UAVs may benefit a lot from the existing researches which focus on dealing with the communication challenges exactly brought by it. In reverse, the flight control also provides a novel method to boost the communication, by changing the positions of the node pair to pursue a LoS link, constructing a stable/regular topology dynamically [265] to reduce the communication overhead, utilizing the depleting UAVs to deliver data for the UAVs along the returning path [54], [55], or letting UAVs flying back and forth as relays to link up two remote areas with a store-carry-and-forward manner [266]. All these works fully exploit the mobility of the UAVs to extend the dimension of the communication decisions, i.e., not only adjusting the communication waveform and protocol parameters, but also altering the geographical locations. So, the vitality of the cyber-physical integration is testified.

TABLE XI
SUMMARY OF THE EXISTING RESEARCHES ON THE UAV NETWORKS FROM A PERSPECTIVE OF THE COUPLING EFFECTS

| Component | Coupling effect | Descriptions | Hierarchy | References |
|---|---|---|---|---|
| *Computation and communication* | Communication contributes to computation | Communication makes global optimization (computation) much easier by building a consensus belief among the individuals. It may also imprison the computation ability, especially for the tasks requiring real-time and high-throughput inputs, so communication improvement will give a positive feedback to the computation ability. | UCL; USL; USoS | [71], [163], [203], [209] |
| | Computation boosts communication | *Intelligent algorithms boost the communication and networking*: The intelligent algorithms could be exploited to improve the performance of the communication and networking from different layers of the network. | USL; USoS | [142], [210]–[225] |
| | | *Multi-agent decisions reduce the communication overhead*: Multi-agent decision can be introduced to let UAVs decide whether to, when to, what to, and whom to communicate. It makes the communication more precise, and the most needed information being transmitted to the most wanted nodes at the most appropriate time can be achieved. | USL; USoS | [205]–[209] |
| *Computation and control* | Computation boosts control | *Computation-enhanced flight control*: The intelligent algorithms can be adopted to design the controller so as to overcome the deficiency of the classic control algorithms. They combine the idea of learning with the conventional control algorithms together, and exploited advantages of the both reasonably to achieve better control effects. | UCL; USL; USoS | [89], [228]–[235] |
| | | *Computation-enhanced formation control*: The formation control of the UAV swarm could be enhanced by adopting the intelligent algorithms, including the bio-inspired algorithms and the artificial intelligent algorithms. | USL; USoS | [170], [237]–[250] |
| *Communication and control* | Communication and control depend on each other | The flight control or the formation control depends on the status interaction among each UAV. In return, the communication network among the UAVs is guaranteed by the optimal topology control, or at least, the proper distance maintenance between each node. | UCL; USL; USoS | [136], [252], [253] |
| | Control-constrained communication | The flight control is mainly dominated by the mission and does not always serve for the communication. Thus, the communication are usually planned and optimized to keep pace with the flight control, and to solve the consequent troubles. | USL; USoS | [22], [23], [52]–[55], [62], [142], [149], [152], [153] |
| | Communication-constrained control | *Formation control under limited communication*: The communication constraints should be considered, including the delays, the limited bandwidth and range, when making the formation control. | USL; USoS | [136], [254], [255] |
| | | *Formation control with communication guarantee*: It means that the communication topology and connectivity among the flying UAVs must been ensured in the formation control. | USL; USoS | [184], [256]–[264] |
| | Promote and benefit from each other | Communication performance improvement will motivate the formation control by efficiently delivering the status among the UAVs. Flight control also creates a novel method to solve the problems in communication by integrating the physical domain, e.g., location changes, other than only making some adjustment to the communication parameters. | USL; USoS | [11], [54], [55], [179], [180], [183], [265]–[268] |

There are many researches concentrating on controlling the mobility or altitude of UAVs to optimize the performance of the UAV network. They can be classified into two categories generally, namely, *i)* incorporating the store-carry-forward capability of the UAVs into the system. Through it, on one hand, exhausting UAVs could bring data to the desiring UAVs when they return back for charging, and thus the system throughput increases [54], [55]. On the other hand, the energetic UAVs could act as mobile relays to link up two remote nodes or clusters, and thus the connectivity is improved [266], [267]. The difference lies in that the latter is to proactively convert the mobility energy into the communication performance, so, there can be a power allocation tradeoff between the flight control and the communication, especially, when the flight and communication share one battery; *ii)* aggregating the dispersive, irregularly distributed and possibly disconnected UAVs to form a connected network with regular topology, such as ring [180], triangular [183], [268] and square [179], or with multi-connection for fault-tolerance [11], [268]. For example, in [11], [268], both works consider the possible communication faults, e.g., caused by energy depletion and malicious attacks, and they aim at building a bi-connected network for fault-tolerance. In practice, bi-connection is desired because UAVs are usually deployed in evil environments.

Table XI summarizes the existing researches on the UAV networks from a perspective of the couplings among communication, computation and control.

## VI. Challenges and Open Issues

As a complex cyber physical system, the UAV network integrates multidisciplinary techniques. Except for all above, there are still many challenges and open issues worthy of attention, from the perspective of either the UAV networks or CPS.

### A. CPS model of the UAV networks

To model the UAV networks from the CPS perspective is challenging. First, the discrete and asynchronous computation processes coexist with the continuous and synchronous physical processes in the UAV networks. Thus, how to realize the tight coupling between the two seemingly paradoxical processes is a fundamental problem in UAV researches [269]. Second, UAV networks involve the interaction of multiple disciplines, including the computation, communication and control, which contains many heterogeneous sub-models. However, there is not a unified theoretical framework that can jointly deal with the computation system, communication network system, control system and the dynamic physical system. Third, the fusion of the computational and physical processes brings the system behaviors and states with more spatio-temporality and dynamic non-determinism. Thus, for the existing modeling methods, most of which are based on the semantic representation, it is difficult to adapt. For example, in the autonomous flight of the UAVs, the computation, communication and control processes continuously interact with the physical environment to determine the behaviors at the next moment. All of them are under the strict spatial-temporal constraints.

CPS modelling of the UAV networks needs to be advanced. First, existing researches on the CPS modelling are all conducted from the event model, agent model or application system model perspective. However, they ignore the influence of time on the system behaviors [270], or simply consider time as a non-functional attribute. Second, many of the existing services, especially the network communication services, adopt the "best effort" ideas, which conflicts with the characteristics of CPS, such as the hard real-time and high reliability [271]. Third, the UAV networks are usually distributed, which makes the CPS components disperse in different locations. Thus, when modelling these distributed systems, the key issues, including the synchronization, network delay and unified system identification, need to be solved [272]. *Edward A. Lee*, the leader of the CPS research, points out six urgent problems in modeling the CPS systems [273], namely, *i)* how to accurately express the model uncertainties, parameter uncertainties and behavior dynamicity of the system; *ii)* how to maintain the consistency of diverse sub-models; *iii)* how to deal with the ambiguity among different CPS components ; *iv)* how to accurately define the time characteristics of the CPS model; *v)* how to deal with the timing sequence inconsistency, network delay, incomplete communication and system state consistency in the distributed CPS model, and *vi)* how to deal with the network and subsystem heterogeneity.

### B. Convergence of heterogeneous networks

CPS usually converge multiple heterogeneous networks. Especially for USoS, it fuses not only multiple different USLs, but also many other existing networks, including the wireless sensor networks, wireless local area networks, cellular networks and so on. UAV networks integrate many techniques and applications of the existing networks, however, they have different characteristics and design objectives compared with the traditional networks, including the dynamicity, heterogeneity, embeddedness and large traffic capacity.

Due to these characteristics, the convergence of multiple networks is of great significance for the further development of the UAV networks. First, each UAV has communication capabilities, and they can build communication network of different levels and scales. Second, due to the mobility and complicated operation circumstance, the communication networks may be partitioned and then aggregated. Third, for a spatial-temporal mission, several heterogeneous UAV swarms may also cooperate. At last, in many applications, the UAV networks will interact with the ground sensor networks, or access the existing network infrastructures (e.g., assisting the terrestrial cellular base stations). Currently, in the researches on modelling the hybrid networks, there are still many challenges that need to be further studied, such as the network node access, channel switching, seamless service handover, network security and quality of service.

In terms of the heterogeneous network protocol, the academia has proposed communication protocol stacks specific for CPS, including CPS2IP and a six-layer communication protocol stack, i.e., CPI. Taking the CPI protocol stack as an example, it inherits the five-layer structure (physical layer, data link layer, network layer, transport layer, and application layer) of the traditional TCP/IP protocol stack, and makes some adjustments according to the characteristics of CPS, such as high real-time demand, flexible structure and so on. A cyber physical layer is added upon the application layer, which is used to describe the characteristics and dynamics of the physical system. However, there are still lots of technical problems to be further studied for the heterogeneous network protocol in CPS/UAV networks.

### C. Computation offloading

Due to the extensive applications of the UAVs, they may encounter various computationally intensive tasks, such as the pattern recognition, natural language processing, video preprocessing and so on. Although the UAVs are equipped with considerable computation power nowadays, it is still not reasonable to leave all these tasks to the UAV themselves. These tasks are typically resource-hungry and high energy consumption, but the UAVs generally have limited computation resources and battery life. Therefore, there will be a dilemma between the resource-hungry applications and the resource-constrained UAVs, which poses a significant challenge for the applications of the UAVs. Computation offloading is envisioned as a promising approach to address such challenges. And it is the core idea of some new emerging computation technologies, such as the mobile cloud computing, mobile edge computation and so on. By offloading some computation tasks via wireless access to the resource-rich infrastructure (i.e., the edge/cloud server, such as the GCS or the intelligent cloud service platform), the capabilities of UAVs for resource-hungry applications could be augmented.

Computation offloading has attracted significant attentions. Many researches consider the computation offloading problem in the scenarios where the UAVs work as the access points and offer computation offloading opportunities to the ground mobile users, for example, in the UAV-enabled wireless powered mobile edge computing system [274], or in the UAV-based mobile cloud computing system [275]. Indeed, these works are inspiring to the computation offloading solutions in the UAV networks where the GCS and the cloud service platform provide computation offload services for the mobile UAVs. However, the mobility of UAVs is much more complicated compared to the ground mobile users, which makes the problem more intractable. There are also some works focusing on the computation offloading in the UAV networks specifically. For example, in [276], the authors propose a new mobile edge computing setup where a UAV is served by the cellular ground base stations for computation offloading. This work tries to minimize the UAV's mission completion time by optimizing its trajectory jointly with the computation offloading scheduling.

In [277], a game theory model is adopted between the UAV players, to solve the computation offloading problem while achieving a tradeoff between the execution time and energy consumption. All in all, computation offloading turns out to be a great help for the resource constrained UAVs. And it deserves to be further developed for the UAV networks with complex missions.

### D. Resource scheduling

Resource scheduling aims at optimizing the resource configurations on the basis of meeting the basic application requirements. There are a great deal of resources in the UAV networks, including the sensing, communication, computation and actuation resources. The system performance of the UAV networks themselves and the quality of service provided to the external can be improved through the efficient resource scheduling and task allocation technologies. There are a few researches on the resource scheduling of the CPS and the UAV networks, and most of them focus on the sensor resource management [278], energy allocation [279], [280], or communication resource scheduling [281].

Actually, there exist many challenges in the UAV resource scheduling, including, i) the resources in the UAV networks are characterized by rich varieties, large quantities and strong heterogeneity, which makes it difficult to establish the resource model. In addition, the research on resource scheduling is still in its infancy, and there are no mature resource management techniques for reference; ii) the evolution process of the physical system is uncertain, and the sensor observations are also interfered by the environment, which bring challenges to the scheduling of the sensing resources; iii) due to the embedded structure, the computation power of single UAV is limited. They can complete complex tasks only through collaboration. Therefore, how to assign tasks to various UAVs to not only meet the interdependence of tasks but also effectively achieve parallel processing is challenging; iv) actuation resource scheduling involves the control to the physical world, which makes the optimizing control and task planning problems more complicated. For example, as a typical actuation resource, mobility is an indispensable for the UAV networks. However, the mission-oriented mobile resource scheduling may face challenges in an uncertain environment.

### E. Energy efficiency

Generally, the energy of the UAVs is quite limited. The whole system life is determined by the amount of the power, in other words, how to efficiently exploit the limited power. In the UAV networks, the communication and motion of the UAVs consume the most part of the energy compared to the sensing and computation. There can be two power supplication cases, i.e., the energy for communication equipment and for powering the UAV comes from the same source, or alternatively, from different sources [20]. In both cases, the energy for communication and control should be tuned well so as to i) respectively prolong the lifespan of communication and flight as much as possible, i.e., improve the energy efficiency for the two parts respectively; ii) jointly extend the whole system life by orchestrating them to achieve a comparable lifetime for the two, since that a power failure for each of the two will deprive the ability of the system to fulfill the mission. There are many researches focusing on improving the energy efficiency of the data transmission to achieve green communication from the perspective of the network layer [282], data link layer [283], physical layer [284] or cross-layer protocols [285], which can be enlightening and applied to the UAV networks. However, the energy efficiency in the UAV networks is much more complicated considering the motion characteristic and their tight coupling with the communication.

Intuitively, the energy efficiency could be improved by fully exploiting the computation power to make optimal decisions on the communication and flight control. For example, the amount of UAVs' transmissions can be reduces by utilizing MDP/POMDP to decide whether/what/whom to communication, choosing the optimum transmission power to satisfy the Qos requirement, or planning the optimal trajectory with lower energy consumption while guaranteeing the mission fulfilment. In addition, as illustrated in Part C, Section V, in the scenario where the shadow effect exists, the improved communication Qos and boosted energy efficiency can still be obtained by moving the UAV for a LoS link, other than purely amplifying the transmission power. Actually, the latter strategy may not work at all and the accessorial energy for communication will be totally wasted. Thus, planning the communication and control jointly based on the coupling of the two will provide a novel solution to improve the energy efficiency. And it needs to be further studied, especially when introducing some artificial intelligent algorithms.

### F. Security and privacy

The closed loop of the UAV networks contains multiple components, i.e., sensing, communication, computation and control, which makes it vulnerable to external attacks since that security problems in any one component will paralyze the whole system. The attacks could be performed from either the physical or the cyber layer. The former means the attacks that directly tamper the physical elements in the UAV networks, e.g., disguising the objectives or changing the batteries. The latter indicates the attacks that are deployed through malware and software, by gaining access to the communication network elements (e.g., jamming or spoofing the UAV transmissions) or by faking the sensor information. To sum up, the major security requirements of the UAV networks include the sensing security, storage security, communication security, actuation control security and the feedback security. In [36], [39], [41], the authors survey the attacks and security issues in the CPS systems, including the smart grids, medical devices and industrial control systems.

Compared to the traditional CPS applications, the security issues in the UAV networks are much more outstanding. First, UAVs always fly in an open space which brings them a complicated and uncontrolled running environment. Second, UAVs communicate through a wireless network, of which the security is much more complicated compared to the wired network. In general, there is a lack of security standards for UAVs, and it has been shown that they are vulnerable to attacks that target either the cyber or physical components [286].

Privacy issues also come with the security problems. The increasing deployment of the UAV networks will result in more perceived data, including the geographical address, and some personally identifiable and sensitive data, for example, when executing a surveillance mission. Such data can be used to profile any individual collectively. Horribly, a physical attack, e.g., on the buildings, humans and installations, can be planned through intruding the storage, communication or actuation modules of the UAV networks. Especially for the USoS, various information from all kinds of UAVs with diverse mission details can be mined from the cloud service platform. Thus, the privacy leakage needs to be considered in the design of a security solution for the UAV networks. In [287], the authors study the security and privacy requirements of the internet of UAVs, and outline potential solutions to address challenging issues, including the privacy leakage, data confidentiality protection and flexible accessibility. In [286], the authors have surveyed the main security, privacy and safety aspects associated with the civilian drone usage in the national airspace. They identify both the physical and cyber threats of such systems, and discuss the security properties required by their critical operation environment. Even so, how to securely and efficiently share these collected data remains an ongoing challenge.

### G. Performance evaluation

What techniques and metrics are chosen to evaluate, verify and optimize the performance of various components is a major challenge in the UAV network design. Although there are plenty of performance indicators and testing platforms for respectively evaluating the communication, computation and control, as well as the physical processes of the UAVs, the evaluation to the CPS characteristics of the UAV networks still lacks theoretical basics and guidelines. To be specific, the CPS model of the UAV networks involves various details of the cyber/physical domain and hardware/software, as well as the heterogeneity of different sub-systems. All these may result in state space expansion, which brings challenges to the system verification and greatly increases the verification cost [288]. On one hand, it is impossible to verify the coupling depth between each component, on the other hand, the contributions of the coupling to the system improvement is also difficult to quantify.

Currently, CPS verification lacks a universally accepted standard. First, we can evaluate the CPS models and methods by referring to the existing modelling simulation and verification techniques. Second, the human-computer interaction systems and the complex system simulation platforms can be fully improved and utilized. For example, the University of Virginia and University of Berkeley have adopted MacroLab as the basic programming and verification environment for CPS and obtained ideal experimental results [289]. Further researches could be conducted from the aspects of the system scheduling capabilities, power consumption, running speed, memory usage, deadlock and privacy.

## H. Unwelcomed social impacts

In spite of the benefits in terms of the military and civilian applications, it must be admitted that the UAVs have brought about some serious risks and unwelcomed impacts to the society, which need to be highlighted. The unwelcomed impacts include but are not limited to public privacy, safety and psychological disturbance [290].

First, deploying UAVs for surveillance purposes by the security sectors primarily aims at preventing the nation and people from the internal and external threats. However, laws regarding the uses of UAVs for military and civilian purposes should be regulated on the national and international level, so as to uphold the basic human privacy rights. Once every act of the citizens is monitored by the state or the bad guys, freedom and human rights will fail to prosper. Technical advancements in these systems should not improve the efficiency of gathering information at the expense of compromising civilian liberties and citizen rights.

Second, UAVs are usually deployed to offer wireless access or autonomous surveillance for temporary events in complex urban environments with large volume of people. These applications are accompanied by potential safety hazards to the public in that UAV crashes may happen due to the power depletion, air turbulence, lightening and unforeseen system errors. In addition, the airspace is going to be shared between the UAVs and other manned aircrafts in urban areas, so, there is also a serious risk of midair collisions which may result in widespread destruction. The whole air traffic control system should be redesigned to deal with the new and continuously variable operational parameters introduced by the integration of the UAVs.

Last but not least, UAVs are expected to provide long range in term of time and space for surveillance applications. The citizens residing in such areas are under immense pressure from being constantly spied by the aerial networks. The psychological impacts of such activities are so profound that are much likely to promote a mentally paralyzed social order. Moreover, the operation of such aerial networks in war-torn areas, such as Syria and Pakistan, has bitterly changed the lifestyle of the citizens. They are suffering through an anxiety of being mistakenly targeted due to the ambiguous information or system errors of the circling UAVs above them. Studies and reports reveal that the civilian communities of these unfortunate areas are in a state of psychological and emotional trauma.

## VII. Conclusion

The UAV networks and CPS are attracting great attention due to their advantages both in extending the range of human activity without being involved. Integrating the UAVs into the CPS system, or developing the UAV networks from a CPS perspective is expected to boost the performance of the UAV networks in executing various complex missions. In this paper, we intend to systematically survey the UAV networks from a CPS perspective. We start the review by presenting some backgrounds and preliminaries with respect to the correspondence between the UAV and CPS, the autonomy level of the UAVs. Further, the main body of the paper is constructed in three aspects: i) the basics of the three cyber components, i.e., communication, computation and control, are respectively discussed, including the requirements, challenges, solutions and advances; ii) the three UAV network hierarchies, i.e., the cell level, the system level and the system of system level, are investigated according to the CPS scales. We explicitly demonstrate the architecture, key techniques and applications of the UAV network under each hierarchy; iii) the coupling effects and mutual inspirations among the UAV network components are excavated, which could be revealing to deal with the challenges from a cross-disciplinary perspective. Finally, we discuss some challenges and open issues with respect to the future researches and the consequent bad effects to the society that need special attention.

It must be admitted that the UAV networks and CPS are both a interdisciplinary and complex, but promising and ongoing area. The related issues, including the requirements, challenges and technologies, can be vast and also emerging. This bounded survey is envisioned to provide a brief tutorial for the beginners, and a novel CPS perspective to deal with the cross-disciplinary challenges for the researchers. We believe that the UAV networks and CPS will flourish by motivating each other, and together make our life more convenient and smarter.

36